\documentclass[twocolumn]{aastex62}
\pdfoutput=1 %for arXiv submission
\usepackage{amsmath,amstext}
\usepackage[T1]{fontenc}
\usepackage{apjfonts} 
\usepackage{multirow}
\usepackage{braket}
\usepackage[figure,figure*]{hypcap}

\usepackage[encapsulated]{CJK}
\usepackage{ucs}
\usepackage[utf8x]{inputenc}

\newcommand{\temp}{$T_{\rm d}$}
\newcommand{\mdust}{$\Sigma_{\rm d}$}

\newcommand{\brat}{$\langle T_{\rm d} \rangle$}
\newcommand{\mbrat}{$\langle T_{\rm d} \rangle_M$}
\newcommand{\lbrat}{$\langle T_{\rm d} \rangle_L$}
\newcommand{\sunit}{$M_\odot$ pc$^{-2}$}

\shorttitle{The Distribution of Dust in Local Group Galaxies}
\shortauthors{Utomo, Chiang et al.}

\begin{document}

\title{THE RESOLVED DISTRIBUTIONS OF DUST MASS AND TEMPERATURE IN LOCAL GROUP GALAXIES}

\author[0000-0003-4161-2639]{Dyas Utomo}
\affiliation{Department of Astronomy, The Ohio State University, 4055 McPherson Laboratory, 140 West 18th Avenue, Columbus, OH 43210, USA}
\author[0000-0003-2551-7148]{I-Da Chiang \begin{CJK*}{UTF8}{bkai}(江宜達)\end{CJK*}}
\affiliation{Center for Astrophysics and Space Sciences, Department of Physics, University of California, San Diego, 9500 Gilman Drive, La Jolla, CA 92093, USA}
\author[0000-0002-2545-1700]{Adam K. Leroy}
\affiliation{Department of Astronomy, The Ohio State University, 4055 McPherson Laboratory, 140 West 18th Avenue, Columbus, OH 43210, USA}
\author[0000-0002-4378-8534]{Karin M. Sandstrom}
\author[0000-0002-5235-5589]{J\'er\'emy Chastenet}
\affiliation{Center for Astrophysics and Space Sciences, Department of Physics, University of California, San Diego, 9500 Gilman Drive, La Jolla, CA 92093, USA}

\correspondingauthor{Dyas Utomo}
\email{utomo.6@osu.edu}

\begin{abstract}
We utilize archival far-infrared maps from the {\it Herschel} Space Observatory in four Local Group galaxies (Small and Large Magellanic Clouds, M31, and M33). We model their Spectral Energy Distribution (SED) from 100 to 500 \micron\ using a single-temperature modified blackbody emission with a fixed emissivity index of $\beta = 1.8$. From the best-fit model, we derive the dust temperature, \temp, and the dust mass surface density, \mdust, at 13 parsec resolution for SMC and LMC, and at 167 parsec resolution for all targets. This measurement allows us to build the distribution of dust mass and luminosity as functions of dust temperature and mass surface density. We compare those distribution functions among galaxies and between regions in a galaxy. We find that LMC has the highest mass-weighted average \temp, while M31 and M33 have the lowest mass-weighted average \temp. Within a galaxy, star forming regions have higher \temp\ and \mdust\ relative to the overall distribution function, due to more intense heating by young stars and higher gas mass surface density. When we degrade the resolutions to mimic distant galaxies, the mass-weighted mean temperature gets warmer as the resolution gets coarser, meaning the temperature derived from unresolved observation is systematically higher than that in highly resolved observation. As an implication, the total dust mass is lower (underestimated) in coarser resolutions. This resolution-dependent effect is more prominent in clumpy star-forming galaxies (SMC, LMC, and M33), and less prominent in more quiescent massive spiral (M31).
\end{abstract}


\section{Introduction}
\label{sec:intro}

Most of the dust mass in galaxies resides in grains that are in thermal equilibrium with the interstellar radiation field \citep[ISRF;][]{draine03}. The strength of the interstellar radiation field is often denoted as $U$, which is the energy density of starlight relative to that measured by \citet{mathis83} for the Solar neighborhood. For a single population of dust in equilibrium with a radiation field, $U$, the dust temperature, \temp, depends on the radiation field in a simple way via $U \propto T_{\rm d}^{4+\beta}$, where $\beta$ is the dust emissivity index. That temperature-ISRF relation assumes the dust cross-section per unit mass, $\kappa$, depends on the wavelength as $\kappa \propto \lambda^{-\beta}$ \citep{draine11}.

Extragalactic observations necessarily convolve together many different environments and radiation fields due to the limited angular resolution of infrared telescopes. To account for this, starting from the work of \citet{dale01}, many extragalactic studies have modelled the infrared spectral energy distribution (SED) using a combination of dust populations, each in equilibrium with a distinct radiation field \citep[e.g.,][]{dale02,draine07b,galliano11,chastenet17}. In this model, a distribution function describes the amount of dust mass, $M_d$, heated by radiation fields, as $dM_{\rm d}/dU$, and is usually modeled as a power law, $U^{-\alpha}$, where $\alpha \sim 2$ \citep{draine07b}.

Some part of this distribution of dust heating, $dM_{\rm d}/dU$, reflects geometry and radiative transfer on very small scales. For example, in the sub-parsec zone of influence of a young stellar population or the photon-dominated region at the edge of a molecular cloud, dust is exposed to a wide range of radiation field intensities \citep[e.g., see][]{dale01}. These effects can be best studied in analytical radiative transfer models, highly resolved observations of Milky Way regions, or via their imprints on the integrated SED.

On larger scales, some part of the dust heating distribution will also arise from variations in the physical conditions across a galaxy \citep[e.g.,][]{gordon14}. The magnitude of these variations can be measured by mapping the resolved temperature and mass of dust across a galaxy. From such maps, we can measure how dust mass is distributed as a function of the illuminating radiation field. Also, we can estimate the effect of blurring together all the regions of a galaxy into a single measurement, which is equivalent to observing a more distant galaxy.

Due to their proximity, Local Group galaxies offer the best opportunity to study the resolved distributions of dust mass and dust temperature across the entire galaxy. In this paper, we use data from the {\it Herschel} Space Observatory \citep{pilbratt10} to derive highly resolved maps ($13-167$~pc resolution) of dust mass surface density and temperature for four Local Group galaxies, the Small and Large Magellanic Clouds (SMC and LMC), M31, and M33. We use these maps to measure how dust mass is distributed as a function of dust temperature, which traces the average illuminating radiation field $U$ in an observational element or a pixel. We compare these distributions among our four targets, and we investigate how degrading the resolution of the data (equivalent to integrating over a range of dust mass and temperatures) would affect the inferred values of the dust mass and temperature at that degraded resolution. We also compare it with the dust temperature and dust total mass derived from integrated SED of unresolved object, that mimics the high-redshift studies \citep[e.g.,][]{magdis10,magdis12,scoville14,genzel15}.

A number of previous papers have utilized the \textit{Herschel} maps of dust emission in each of our targets galaxies individually. These include the studies of the SMC and LMC by \citet[][]{meixner13,gordon14,duval14,chastenet17}, the M31 work by \citet[][]{fritz12,groves12,smith12,draine14}, and the M33-focused investigations of \citet[][]{braine10,boquien11,xilouris12}. The new contributions of this paper include homogenizing the methodology and analysis for all targets, focusing on implications of the small scale structures ($\sim 10-100$~pc) for observations of more distant galaxies, and employing the new fitting code developed by \citet{chiang18}, based on \citet{gordon14}.

Throughout this paper, we use the five longest wavelengths of {\it Herschel} bands (100~\micron, 160~\micron, 250~\micron, 350~\micron, and 500~\micron). We expect the infrared emission at these wavelengths mainly captures emission from relatively large grains in thermal equilibrium with the local radiation field. These grains tend to represent the dominant mass component, and focusing on this wavelength range allows us to employ a modified blackbody model to fit the SED, following \citet{chiang18} and \citet{gordon14} methodologies.

This paper is organized as follows. We describe the archival data in $\S$\ref{sec:two} and our modified blackbody modeling in $\S$\ref{sec:mbb}. We present maps of dust mass surface density and temperature for all four targets in $\S$\ref{sec:three}. We also show the distribution of dust mass as functions of illuminating radiation field for each galaxy  and locations within a galaxy in $\S$\ref{sec:three}. We show the correlation between dust mass surface density vs. dust temperature, IR color vs. dust temperature, and 500 \micron\ intensity vs. dust mass surface density in $\S$\ref{sec:corr}. We explore how the inferred dust temperature changes as a function of resolution (down to treating the galaxy as one unresolved source) in $\S$\ref{sec:reso}. Finally, we summarize our findings in $\S$\ref{sec:sum}.

Throughout this work, we adopt a distance of 62.1~kpc to the SMC \citep{graczyk14}, 50.2~kpc to the LMC \citep{klein14}, 744~kpc to M31 \citep{vilardell10}, and 840~kpc to M33 \citep{freedman91}. We also adopt inclination of $2^{\circ}.6$ for SMC \citep{subramanian12}, $34^{\circ}.7$ for LMC \citep{vanmarel01}, $77^{\circ}.7$ for M31 \citep{corbelli10}, and $56^{\circ}$ for M33 \citep{paturel03}. We summarize the symbols used in this paper in Table~\ref{tab:index}.

\begin{deluxetable*}{ c c c c }
\tablewidth{0pt}
\tabletypesize{\scriptsize}
\tablecaption{Descriptions of symbols used in this paper}
\label{tab:index}
\tablehead{
\\
Symbols & Descriptions & Measured for & References
}
\startdata
\multicolumn{4}{c}{Temperature [K]} \\
\hline
\temp\ & Expectation value of the dust equilibrium temperature at a given map resolution & a pixel & Equation~\ref{eq:fit_temp} \\
$T_{d,L}$ & Luminosity-weighted of \temp\ at a given map resolution & a pixel & Equation~\ref{eq:tdl} \\
\mbrat\ & Mass-weighted mean of \temp\ from a given map resolution & whole galaxy & Equation~\ref{eq:mwtemp} \\
\lbrat\ & Luminosity-weighted mean of \temp\ from a given map resolution & whole galaxy & Equation~\ref{eq:lwtemp} \\
%$T_{\rm int}$ & Dust temperature from integrated SED using single- or multi-temperature model & whole galaxy & Section~\ref{sec:int_fit} \\
$T_{\rm peak}$ & Dust temperature that corresponds to the peak of mass or luminosity distribution & distribution function & Section~\ref{sec:dis_mass_by_u} \\
$T_{\rm med}$ & The median value of dust temperature in the mass or luminosity distribution & distribution function & Section~\ref{sec:dis_mass_by_u} \\
$\mathcal{W}_T$ & The 16th-to-84th percentile range of temperature in mass or luminosity distribution & distribution function & Section~\ref{sec:dis_mass_by_u} \\
\hline
\multicolumn{4}{c}{Mass [$M_\odot$] and Mass Surface Density [$M_\odot$pc$^{-2}$]} \\
\hline
$\Sigma_d$ & Expectation value of the dust mass surface density at a given map resolution & a pixel & Equation~\ref{eq:fit_mass} \\
$\Sigma_{d,L}$ & Luminosity-weighted of $\Sigma_d$ at a given map resolution & a pixel & Equation~\ref{eq:sdl} \\
$M_{\rm d,Tot}$ & Total mass from integrated SED using single- or multi-temperature model & whole galaxy & Section~\ref{sec:int_fit} \\
$\Sigma_{\rm peak}$ & Dust mass surface density that corresponds to the peak of mass or luminosity distribution & distribution function & Section~\ref{sec:dis_mass_by_mass} \\
$\Sigma_{\rm med}$ & The median value of dust mass surface density in the mass or luminosity distribution & distribution function & Section~\ref{sec:dis_mass_by_mass} \\
$\mathcal{W}_\Sigma$ & The 16th-to-84th percentile range of mass surface density in mass or luminosity distribution & distribution function & Section~\ref{sec:dis_mass_by_mass} \\
\hline
\multicolumn{4}{c}{Interstellar Radiation Field} \\
\hline
$U$ & Interstellar radiation field strength (energy density) at a given resolution relative to Solar neighborhood value & a pixel & Equation~\ref{eq:u2t} \\
$U_{\rm min}$ & The minimum value of $U$ used in multi-temperature modeling & model & Equation~\ref{eq:distU} \\
$U_{\rm max}$ & The maximum value of $U$ used in multi-temperature modeling & model & Equation~\ref{eq:distU} \\
$U_{\rm peak}$ & Interstellar radiation field strength that corresponds to the peak of mass or luminosity distribution & distribution function & Section~\ref{sec:dis_mass_by_u} \\
$U_{\rm med}$ & The median value of $U$ in the mass or luminosity distribution & distribution function & Section~\ref{sec:dis_mass_by_u} \\
$\mathcal{W}_U$ & The 16th-to-84th percentile range of $U$ in mass or luminosity distribution & distribution function & Section~\ref{sec:dis_mass_by_u} \\
\hline
\multicolumn{4}{c}{Others} \\
\hline
$\mathcal{J}_{\rm IR}$ & A quantity proportional to the IR equilibrium luminosity at a given map resolution & a pixel & Equation~\ref{eq:tir} \\
$\alpha$ & The slope of the power-law distribution of dust mass heated by $U$ & model & \citet{dale01} \\
$\beta$ & The dust emissivity index (power law exponent in the dependency of $\kappa$ as a function of wavelength) & model & Equation~\ref{eq:beta} \\
$\gamma$ & The fraction of dust mass heated by a power law distribution field, $U^{-\alpha}$, between $U_{\rm min}$ and $U_{\rm max}$ & model & \citet{draine07a} \\
$\kappa$ & Dust absorption cross-section & model & Equation~\ref{eq:opt_depth} \\
$\mathcal{L}$ & Likelihood of a given model & model & Equation~\ref{eq:likelihood} \\
$r_s$ & Spearman rank correlation coefficient & data & \citet{spearman04}
\enddata
\tablecomments{We always specify which resolution, which distribution (mass or luminosity), and which model (single- or multi-temperature) in the text whenever it is necessary.}
\end{deluxetable*}

\section{Data} \label{sec:two}

\subsection{Herschel Far-infrared Maps}

We utilize the far-infrared maps at $\lambda = 100-500$ \micron\ as observed by the {\it Herschel} Space Observatory \citep{pilbratt10} using the PACS \citep{poglitsch10} and SPIRE \citep{griffin10} instruments. These maps were the data products of several key projects: HERITAGE used \textit{Herschel} to observe the SMC and LMC \citep{meixner13,gordon14,duval14}, \citet{groves12} and \citet{draine14} present \textit{Herschel} observations of M31 \citep[see also][]{smith12}, and HerM33s \citep{braine10,boquien11,xilouris12} used \textit{Herschel} to observe M33.

Before any analysis, we convolve each map to convert the point spread function (PSF) from the original PSF of the instrument to a symmetric Gaussian PSF. We use the kernel that is appropriate for each PACS and SPIRE band \citep{aniano11} to do this conversion. This Gaussian PSF allows easy convolution to any lower spatial resolution using coarser Gaussian kernels. After PSF conversion, these PACS and SPIRE maps have full width at half max (FWHM) resolutions of 15\arcsec\ for 100 \micron\ and 160 \micron, 30\arcsec\ for 250 \micron\ and 350 \micron, and 41\arcsec\ for 500 \micron. We refer to these values as the ``native resolutions.''

\subsection{Region of Interest} 
\label{sec:masking}

We identify a region of interest for each galaxy based on the SPIRE maps. We use the SPIRE maps to identify the region of interest because these maps have the best sensitivity to low dust column densities and cool dust. This region represents our estimate of the full spatial extent of the galaxy as detected by \textit{Herschel}. We avoid this region of interest when fitting foreground and background emission, and we only fit the resolved SED within this region of interest.

To identify the region of interest, we first calculate the median value away from the galaxy and median absolute deviation (MAD) in the 500 \micron\ map. Then, we define a threshold for significant emission equal to the median value plus $3 \times$ the noise level as inferred from the median absolute deviation (MAD, Table~\ref{tab:thresh}). We consider the contiguous region of pixels with intensity above this threshold that is also contiguous with the main body of the galaxy to be our region of interest. We further dilate the region of interest by several beam widths after applying the threshold. This dilation includes faint emission around the galaxy and remove any holes in the region of interest.

These regions of interest do an excellent job of covering the whole body of each galaxy (shown as white contours in Figure~\ref{fig:maps_temp}). Given this, the details of this masking process have little impact on our overall results. We verified this by changing our adopted threshold to use a signal to noise cut of $2$ or $4$ and also changing the number of iterations to grow the mask. Modest changes only slightly affect the low-end tail of the distribution of dust mass surface density. The total dust mass in LMC, M31, and M33 only varies within few percent by using smaller or bigger mask. The region of interest definition has more effect on SMC, which is surrounded by extended, faint tidal features. As a result, changing the region of interest in the SMC can cause the total mass in our analysis to vary by $10-20\%$. As Figure~\ref{fig:maps_temp} shows, our adopted region of interest does a good job of including the main body of the galaxy and the bright regions of the SMC's eastern wing.

\begin{deluxetable}{ c | c c c c }
\tablewidth{0pt}
\tabletypesize{\scriptsize}
\tablecaption{The median value and the median absolute deviation (MAD) of the 500~\micron\ map. \label{tab:thresh}}
\tablehead{
\\
{} & SMC & LMC & M31 & M33
}
\startdata
Median [MJy sr$^{-1}$] & 0.012 & 0.187 & 1.663 & 0.135 \\
MAD [MJy sr$^{-1}$] & 0.215 & 0.493 & 0.212 & 0.210 \\
\enddata
\end{deluxetable}

\subsection{Foreground and Background Subtraction} \label{sec:bg_sub}

Each \textit{Herschel} image blends emission from the galaxy with foreground and background emission. These foreground and background contributions reflect a mixture of Milky Way cirrus, the cosmic infrared background and resolved background galaxies, and the adopted observing and imaging strategies which can result in zero-point offsets and gradients. We correct each map for this contamination in several steps.

First, following \citet{bot04}, we use a Milky Way 21-cm map as a template to correct the SMC and LMC maps for foreground emission from the Galactic dust. Because of the wide sky coverage of these two galaxies, using Galactic 21-cm map represents the best way to capture foreground variations across those galaxies. We use the foregrounds calculated by \citet{chastenet17} using the {\sc Hi} maps from \citet{stanimirovic99} and \citet{staveley-smith03}. We refer to \citet[][]{chastenet17} for more details.

After the first step, for all galaxies and all bands, we estimate the combined foreground and backgrounds emission (hereafter just ``background'') and subtract it from the map. To do so, we calculate the median intensity of the image outside the region of interest. We subtract this value from the whole image. We further refined this background estimate by fitting a plane outside the region of interest. We iterate this fit, and dropping pixels that deviated by more than $\pm 2\sigma$ from the fit, where $\sigma$ is the robustly estimated RMS noise. Exclusion of these outlier pixels is important to exclude pixels that may be contaminated by instrumental artifacts or non-background emission. Then, we subtract the fit plane from the whole image. Finally, we construct a histogram of intensities outside the region of interest and make one final adjustment to the zero point of the image, forcing it to coincide with the mode of the histogram. This final step usually represent a small adjustment (significantly less than $1\sigma$).

After this step, the zero point of all bands and all galaxies outside the region of interest appears reasonably consistent with zero intensity.

\subsection{Convolution and Reprojection}\label{sec:convolution}

We convolve the background-subtraced maps from their native angular resolutions to a set of common physical resolutions at the adopted distance of each galaxy.  For the SMC and the LMC, we create maps with FWHM from 13 to 500~pc resolution. For M31 and M33 we create maps with FWHM from 167 to 2,000~pc resolution. In the following analysis, we work mostly at the finest physical scale, i.e. 13~pc for SMC and LMC, and 167~pc for all targets. We choose this scale to match the angular resolutions of SPIRE at 500~\micron\ at the distance to the SMC and M33. Thus, the finest common resolution to study all four galaxies, while still including all SPIRE bands, is 167~pc.

After convolution, we reproject the images from all bands onto a common coordinate grid. We choose the pixel spacing for the common grid so that there are 2.5 pixels across each (FWHM) beam, i.e. we oversample at roughly the Nyquist sampling rate. This means the pixel size in parsec is also larger in coarser resolution. To carry out the reprojection and smoothing, we use the {\tt CASA} tasks {\tt imsmooth} and {\tt imregrid} \citep{mcmullin07}.

\section{Modeling} \label{sec:mbb}

\subsection{Modified Blackbody Model}

The intensity of dust emission, $I_\nu$, for dust in equilibrium with the radiation field depends on the dust optical depth, $\tau (\lambda )$, and the dust equilibrium temperature, \temp, via

\begin{equation}
\label{eq:rad_trans}
I_\nu (\lambda) = B_\nu(T_{\rm d}, \lambda)~[1 - e^{-\tau}],
\end{equation}
where $B_\nu(T_{\rm d},\lambda)$ is the Planck function for temperature \temp\ at a wavelength $\lambda$. 

For our target galaxies and resolutions, dust emission at $\lambda \geq 100~\micron$ is almost always optically thin. In this case,  

\begin{equation}
\label{eq:opt_thin}
I_\nu (\lambda ) \approx \tau (\lambda) ~B_\nu(T_{\rm d}, \lambda)~.
\end{equation}

The dust optical depth can be written as
\begin{equation}
\label{eq:opt_depth}
\tau (\lambda) = \kappa (\lambda)~\Sigma_{\rm d}~,
\end{equation}
where $\kappa (\lambda$) is the dust absorption cross-section per unit mass at a wavelength $\lambda$, and $\Sigma_{\rm d}$ is the dust mass surface density. 

Fitting the SED yields $\tau (\lambda)$, but for a given $\tau (\lambda )$, the values of $\kappa (\lambda)$ and $\Sigma_{\rm d}$ are degenerate. Breaking this degeneracy requires an independent calibration of $\kappa (\lambda )$. Such a calibration, in turn, requires that $\tau (\lambda)$ be measured in a location where $\Sigma_{\rm d}$ is known from independent measurements. The most common site for such calibrations is the local diffuse ISM in the Milky Way, where the spectroscopic measurements of depletion into dust yield an independent estimate of the dust mass surface density \citep[e.g.,][]{jenkins09}, while the SED has been measured by all-sky IR mapping missions. 

We follow this approach, by adopting the calibration scheme of \citet{chiang18}, based on \citet{gordon14}. We calculate $\tau (\lambda )$ by fitting the Solar Neighborhood cirrus emission using our modified blackbody model. Then, we derive $\kappa (\lambda)$ from $\tau (\lambda )$ by adopting the dust abundance found by \citet{jenkins09}. We further assume that $\kappa (\lambda)$ remains constant across all regions in Local Group galaxies. Though variations in $\kappa$ do represent a systematic uncertainty, this procedure removes another systematic uncertainty by using the same dust model to estimate $\kappa$ and fit the data.

For dust, the absorption cross section per unit mass decreases with increasing wavelength. We follow standard practice and assumed it to be a power law \citep{draine11} as 
\begin{equation}
\label{eq:beta}
\kappa (\lambda ) = \kappa ({\lambda_0}) \left(\frac{\lambda}{\lambda_0}\right)^{-\beta}~,
\end{equation}
where $\lambda_0$ is a reference wavelength and $\beta$ represents the dust emissivity index. In this paper, we assume a fixed $\beta = 1.8$ \citep{dunne01,draine07b,clements10,planck11,scoville14}. According to \citet{chiang18}, the variation in $\Sigma_d$ from different choices of $\beta$ values would be small since $\kappa(\lambda_0)$ is calibrated accordingly.

Combining those assumptions above, we get
\begin{equation}
\label{eq:mbb}
I_\nu (\lambda) = \kappa ({\lambda_0}) \ \Sigma_{\rm d} \left(\frac{\lambda}{\lambda_0}\right)^{-\beta} B_\nu(T_{\rm d}, \lambda)~.
\end{equation}
We adopt $\lambda_0 = 160~\micron$ as our reference wavelength. For $\beta = 1.8$, our fit to the Milky Way cirrus yields $\kappa(\lambda_0) = 18.7 \pm 0.6$~cm$^2$~g$^{-1}$.

\subsection{Fitting the Dust SED} \label{sec:fitting}

For each line-of-sight, we estimate $\Sigma_{\rm d}$ and $T_{\rm d}$ by simultaneously fitting the intensities at 100 \micron, 160 \micron, 250 \micron, 350 \micron, and 500 \micron\ using the model in Equation~\ref{eq:mbb}. We follow the algorithms presented in \citet{chiang18} and \citet{gordon14} to calculate the relative likelihood of the model, $\mathcal{L}$, given the observed SED. Here,
\begin{equation}
\label{eq:likelihood}
\mathcal{L} = {\rm exp}(-\chi^2/2)~,
\end{equation}
and 
\begin{equation}
\label{eq:chi2}
\chi^2 = \Delta I^T \mathcal{C}^{-1} \Delta I~,
\end{equation}
where $\Delta I$ indicates a vector recording the difference between the observed SED and the model SED. Before calculating $\Delta I$, we integrate the models (Equation~\ref{eq:mbb}) with the appropriate response function of that \textit{Herschel} band.

$\mathcal{C}^{-1}$ refers to the the inverse of the covariance matrix. This term takes into account the inter-band covariances in the uncertainty calculation. $^T$ denotes matrix transpose operation. In the \citet{gordon14} approach (upon which \citet{chiang18} is based), this covariance matrix, $\mathcal{C}$, is calculated by summing the covariance matrix from the data outside the region of interest and the covariance matrix of the calibration uncertainty from the \textit{Herschel} instruments \citep{balog14,bendo17}. The covariance of the data outside the region of interest are taken to indicate the covariance in the observational noise.

The fitting process calculates the likelihood, $\mathcal{L}$, for each model across a grid of $T_{\rm d}$ and $\Sigma_{\rm d}$. The grid for $T_{\rm d}$ spans from 5 to 50~K with increments of 0.5~K. We space the grid for $\Sigma_{\rm d}$ logarithmically, covering from $-4$ to $1$~dex in log$_{10}\Sigma_{\rm d}$ with a step size of 0.025~dex.

From the grid of likelihoods, we calculate the expectation values for $T_{\rm d}$ and $\Sigma_{\rm d}$ via
\begin{equation}
\label{eq:fit_temp}
T_{\rm d} = \frac{\sum_i~\mathcal{L}_i~T_{{\rm d},i}}{\sum_i~\mathcal{L}_i}~,
\end{equation}
and
\begin{equation}
\label{eq:fit_mass}
\log_{10}\Sigma_{\rm d} = \frac{\sum_i~\mathcal{L}_i~\log_{10}\Sigma_{{\rm d},i}}{\sum_i~\mathcal{L}_i}~.
\end{equation}
Here, the sum, $\sum_i$, covers all cells in the grid. We adopt these expectation values as our estimate of the best-fit $\Sigma_{\rm d}$ and $T_{\rm d}$ at each pixel inside the region of interest. We correct $\Sigma_d$ for galaxy inclination, $i$, by multiplying $\Sigma_d$ with cos($i$).

We repeat this fit and the covariance estimation, at several resolutions. For this paper the key resolutions are $13$~pc (the finest common resolution achievable for both the SMC and LMC) and $167$~pc (the finest common resolution achievable for all four galaxies).

\subsection{Interstellar Radiation Field Strength and Equilibrium Dust Luminosity}
\label{sec:isrf}

For dust in thermal equilibrium with the local radiation field and with a power-law mass absorption coefficient $\kappa$, \temp\ relates to the interstellar radiation field (ISRF), $U$, via
\begin{equation}
\label{eq:u2t}
{\rm log}_{10}~U = (4 + \beta) \ {\rm log}_{10}\left(\frac{T_{\rm d}}{18 \ {\rm K}}\right)~.
\end{equation}
The normalization takes $T_{\rm d} = 18$~K to corresponds to the Solar Neighborhood radiation field, $U_\odot = 1$ \citep{draine14}. This normalization agrees well with our fit to calibrate $\kappa$. In that case, we found $T_{\rm d} = 18.29 \pm 0.11$~K for the local cirrus. Assuming that $U_\odot$ heats local cirrus, then this supports equating $T_{\rm d} \approx 18$~K with $U \approx 1$.

The infrared luminosity emerging from grains that are in equilibrium with the local radiation field is the integral of Equation~\ref{eq:opt_thin} over infrared wavelengths. For simplicity, we define a quantity proportional to the infrared luminosity per unit area as
\begin{equation}
\label{eq:tir}
\mathcal{J}_{\rm IR} \equiv \Sigma_{\rm d}~T_{\rm d}^{4+\beta}~.
\end{equation}
Here, $\mathcal{J}_{\rm IR}$ represents the infrared luminosity surface density of grains with mass surface density $\Sigma_{\rm d}$ at temperature $T_{\rm d}$. We refer to $\mathcal{J}_{\rm IR}$ as the ``equilibrium luminosity surface density.''

Even though the infrared luminosity captures the integral of infrared SED under our best-fit model, it underestimates the true total infrared luminosity because we do not fit or consider emission from hot grains and/or small grains that are out of equilibrium with the local radiation field. Instead, it captures the total light emerging from the large grain population in equilibrium with the local ISRF. Contributions from these hot grains and small grains that are out of equilibrium only make a small difference in the total dust mass. However, they produce an important fraction of the total infrared luminosity emitted by all kind of dust.

\subsection{Caveats}

We adopt several simplifying assumptions that the reader should bear in mind. We assume a single \temp\ for each pixel. In reality, \temp\ may vary within any given resolution element, especially in star-forming regions, which will contain many local heating sources. Hence, our \temp\ measured at 13 or 167~pc resolution is a luminosity weighted mean of the sub-resolution \temp\ distribution. Our paper addresses the large-scale averaging that depends mainly on galaxy structure and the location of star-forming regions. The properties of the small scale distribution of \temp\ \citep{draine07a} are beyond the scope of our work, and will be addressed in upcoming paper (Chastenet et al. in preparation). We refer the reader to \citet{bianchi13} who performed a comparison between the single-temperature model and the full dust distribution model.

Variations in $\beta$ also affect the best-fit \temp, with the sense that a higher value of $\beta$ gives a lower value of the best-fit \temp\ \citep{kelly12}. There are some measurements suggesting a variation of $\beta$ within and among Local Group galaxies \citep[e.g.,][]{smith12,gordon14}. We fix $\beta$ in the interests of simplicity and to avoid the ambiguities in the interpretation, as highlighted by \citet{kelly12}, but this remains an important open topic.

For the dust mass, our measurements depend on the calibrated value of $\kappa(\lambda_0) = 18.7 \pm 0.6$~cm$^2$~g$^{-1}$ for $\lambda_0 = 160~\micron$ and $\beta = 1.8$. This value relies on the assumption that the optical properties of dust grains throughout the Local Group galaxies is the same as in the Milky Way cirrus. If $\kappa(\lambda_0)$ is underestimated, we will overestimate the dust mass, and vice-versa. In order to partially control for this uncertainty, we often normalize our results by the total dust mass in each galaxy. In the case that $\kappa(\lambda_0)$ varies from galaxy-to-galaxy, but not between regions within a galaxy, our results will still be robust.

\section{Maps and Distributions of Dust Mass, Luminosity, and Temperature} \label{sec:three}

\begin{figure*}
\centering
\includegraphics[width=0.95\textwidth]{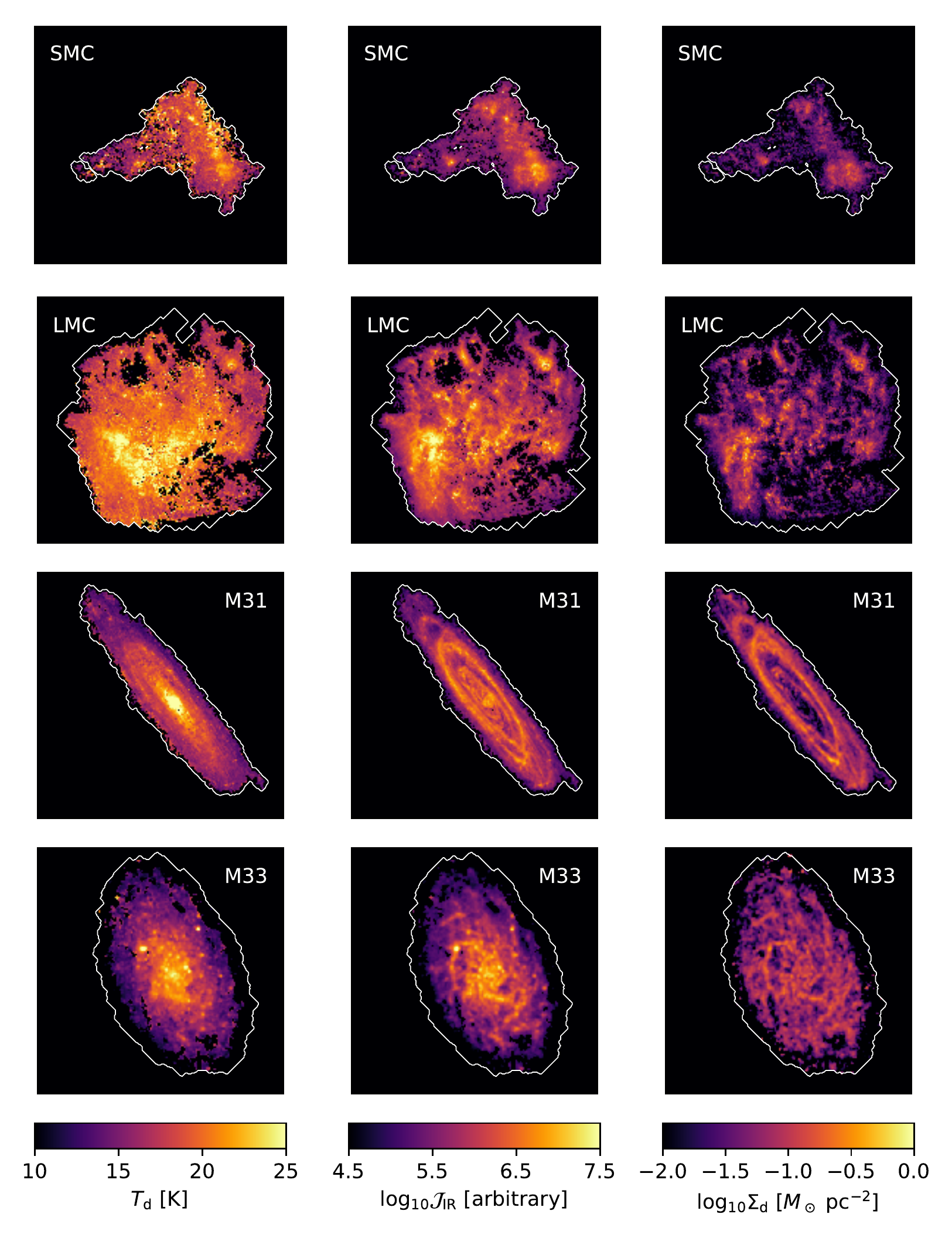}
\caption{\textbf{Dust Temperature, Luminosity, and Mass Surface Density for Local Group Galaxies.} Maps of dust temperature (left panels), equilibrium IR-luminosity (middle panels), and dust mass surface density (right panels) at 13~pc resolution for both LMC and SMC, and at 167~pc resolution for both M31 and M33. White contours mark the region of interest ($\S$\ref{sec:masking}). In the LMC, SMC, and M33, high dust temperatures appear associated with star-forming regions. In M31, the old stellar populations heats the dust in the bulge to higher temperatures \citep{groves12,draine14,viaene14}. In all targets, the dust surface density maps visually resemble other maps of the ISM, especially atomic hydrogen \citep[see][]{kim98,stanimirovic99,staveley-smith03,braun09,braun12,koch18}.}
\label{fig:maps_temp}
\end{figure*}

Figure~\ref{fig:maps_temp} shows our best estimates of dust equilibrium temperature, \temp, equilibrium IR luminosity surface density, $\mathcal{J_{\rm IR}}$, and dust mass surface density, $\Sigma_{\rm d}$, for each target.\footnote{These maps are available in {\tt FITS} file at \url{https://www.asc.ohio-state.edu/astronomy/dustmaps/}} In the SMC, LMC, and M33, we find high dust temperatures in regions of active star formation, consistent with previous works \citep[e.g.,][]{xilouris12,gordon14}. In M31, the highest dust temperatures coincide with the low dust mass surface density in the inner, bulge-dominated part of the galaxy. Previous work has demonstrated that this hot dust in the inner part of M31 results from heating by the old stellar population \citep[][]{groves12,draine14,viaene14}.

The dust mass surface density maps show the same features as gas mass surface density maps in these galaxies. In the SMC, the ISM material is concentrated along the bar, with an extension to the east into the wing, hosting the bright star-forming complex N83/N84. The LMC shows a prominent ridge of material south of 30 Doradus along the eastern edge of the galaxy and shells through the rest of the galaxy. The dust in M31 is concentrated into a series of rings that may be tightly wound spirals \citep[e.g.,][]{nieten06,gordon06}. And in M33, the dust mass maps show flocculent spiral structures.

Atomic hydrogen, {\sc Hi}, makes up most of the neutral ISM in each of our targets. As a result, our $\Sigma_{\rm d}$ maps resemble {\sc Hi} 21-cm maps of these galaxies \cite[see][]{kim98,stanimirovic99,staveley-smith03,braun09,braun12,koch18}. In detail, our maps should also reflect the presence of molecular gas and variations of the dust-to-gas ratio. Even accounting for dust-to-gas ratio variations, Figure \ref{fig:maps_temp} gives among the most uniform views of the ISM in Local Group galaxies up to date.

The luminosity maps ($\mathcal{J_{\rm IR}} \equiv \Sigma_d T_d^{5.8}$) show the brightest regions where most infrared light comes from. These bright regions are due to high $\Sigma_d$, high $T_d$, or both  (excluding emission from hot, small grains). The star forming regions are still bright because they have both warm \temp\ and dense $\Sigma_d$. M31 shows distinctive feature, where the bulge (that shows a `hole' in the $\Sigma_d$ map) is luminous because of its high \temp.

Based on these maps, we calculate the distributions of dust mass and luminosity as functions of local illuminating radiation field, $U \propto T_d^{5.8}$ (in $\S$\ref{sec:dis_mass_by_u}), and local dust mass surface density, $\Sigma_{\rm d}$ (in $\S$\ref{sec:dis_mass_by_mass}). We compare those distributions in star-forming regions against the rest of area in galaxies in $\S$\ref{sec:reg}.

\subsection{Distributions of Dust Mass and Luminosity as a Function of Illuminating Radiation Field}
\label{sec:dis_mass_by_u}

\begin{figure*}
\centering
\epsscale{1.2}
\plotone{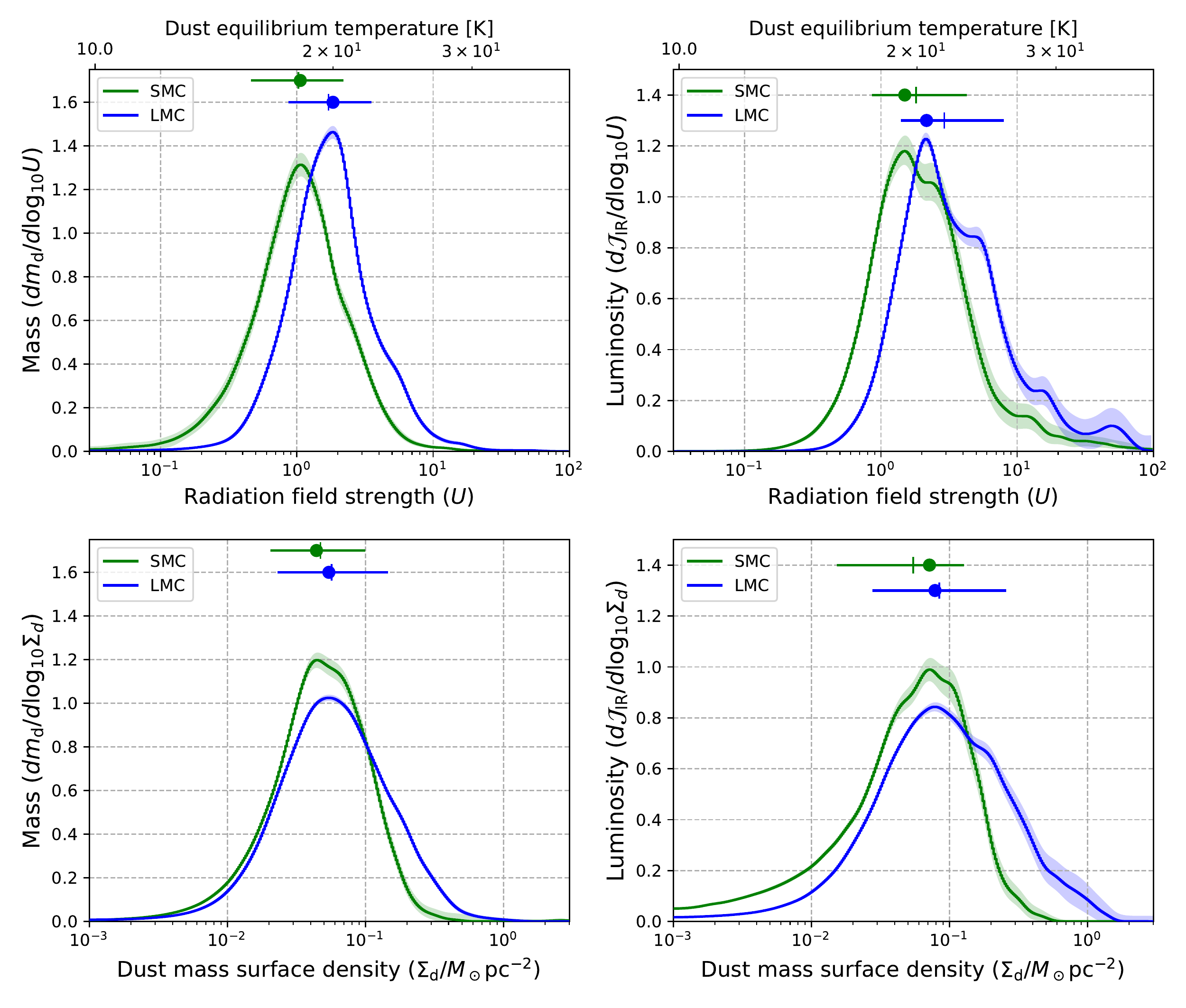}
\caption{{\bf Distributions of dust mass (left) and equilibrium IR luminosity (right) at 13~pc resolution for the Small and Large Magellanic Clouds} as functions of illuminating radiation field, $U$ (top panels), and dust mass surface density, $\Sigma_d$ (bottom panels). We calculate the distributions in $0.01$~dex wide bins, normalize it to the total mass or total luminosity in the galaxy, and then smooth the distribution using a five bin-wide Gaussian filter. Shaded regions indicate $16-84^{\rm th}$ percentile confidence. Circles, ticks, and lines at the top of the plot are the peak, median, and 16$^{\rm th}$-to-84$^{\rm th}$ percentile width of the distribution. Characteristic widths for all four distributions are $\sim 0.6{-}0.95$~dex, but the detailed distributions differ. The luminosity distributions tend to reflect higher mass surface density, higher radiation field regions than that in the mass distribution. Both galaxies show significant contributions from individual bright regions to the luminosity distribution; the ``bumps'' in the distribution reflect the bight spots in the luminosity maps (middle column) of Figure \ref{fig:maps_temp}. We quantify these distributions in Table \ref{tab:params}. We compare the Magellanic Clouds to M31 and M33 at a common resolution of $167$~pc in Figure \ref{fig:dist_167pc}.}
\label{fig:dist_13pc}
\end{figure*}

\begin{figure*}
\centering
\epsscale{1.2}
\plotone{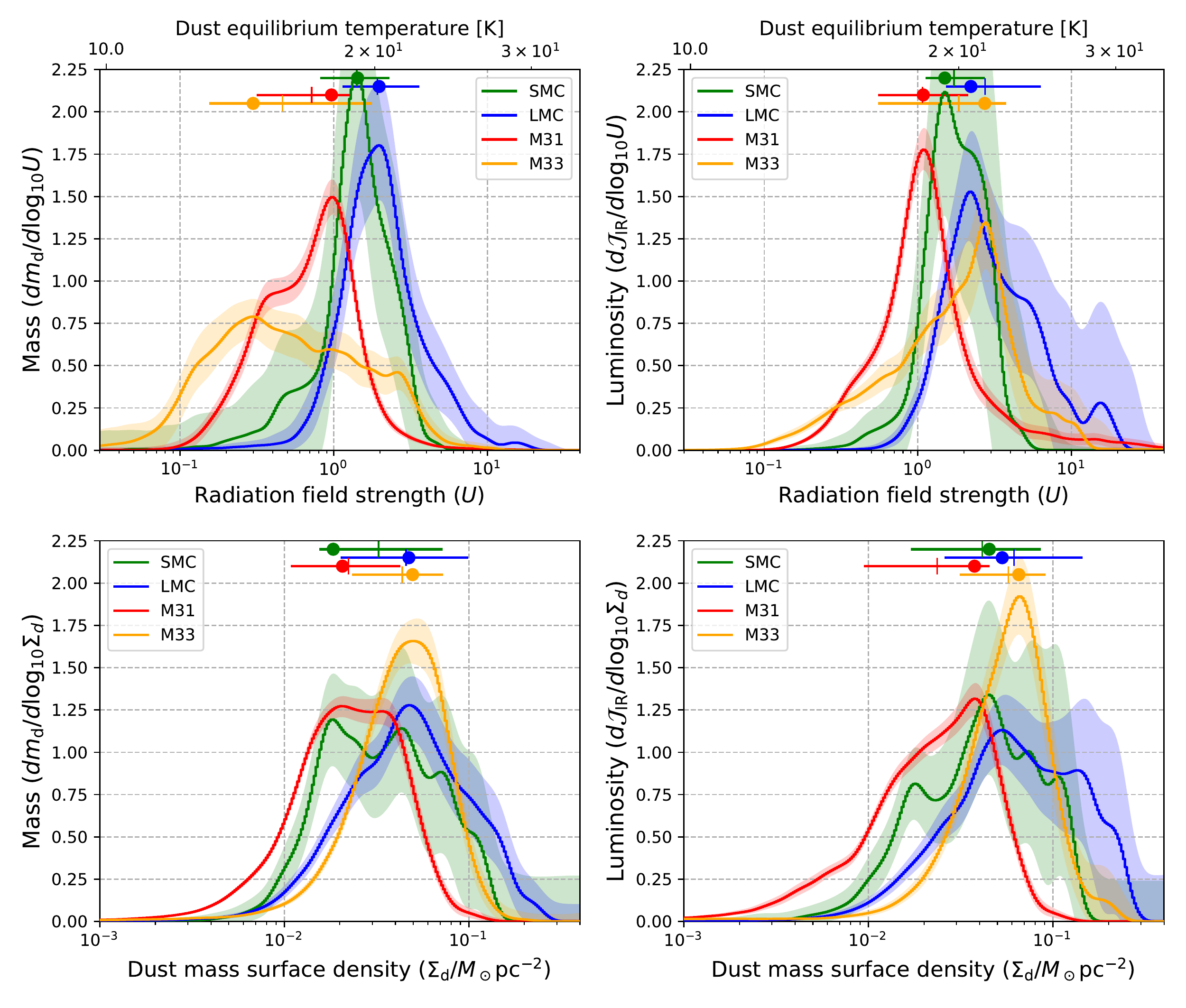}
\caption{\textbf{Distributions of dust mass (left) and equilibrium IR luminosity (right) at 167~pc resolution for all four targets.} The plots show distributions of dust mass (left) and equilibrium IR luminosity (right) as a function of (top) illuminating radiation field, $U$, and (bottom) dust mass surface density, $\Sigma_d$. We calculate the distributions in $0.01$~dex wide bins, normalize to the total mass or luminosity in the galaxy, and then smooth the distribution using a five bin-wide Gaussian filter. Shaded regions indicate $16-84^{\rm th}$ percentile confidence; for the SMC and LMC these are larger than in Figure \ref{fig:dist_13pc} because of the smaller number of pixels at this resolution. Circles, ticks, and lines at the top of the plot the peak, median, and 16$^{\rm th}$-to-84$^{\rm th}$ percentile width of the distribution. The majority of dust mass and luminosity in Magellanic clouds originates from regions with ISRF higher than the Solar neighborhood value (log$_{10}U_\odot$ = 0). In M31 and M33, most of dust mass has $U<U_\odot$, but this is different in the luminosity distribution, where most luminosity in M31 at around $U_\odot$ and higher than $U_\odot$ in M33.}
\label{fig:dist_167pc}
\end{figure*}

The dust mass distribution as a function of illuminating interstellar radiation field, $U$, strongly affects the integrated SED of a galaxy. Though we cannot measure the sub-resolution scale distribution (e.g., photo-dissociation region or {\sc Hii} region structure), our maps allow us to measure the dust mass distribution as a function of $U$ on galactic scales. We measure and plot these distributions in the top panels of Figure~\ref{fig:dist_13pc} (for the Magellanic Clouds at 13~pc resolution) and the top panels of Figure~\ref{fig:dist_167pc} (for all targets at $167$~pc resolution).

Following \citet{dale01}, we write these distributions as $dM_{\rm d}/d{\rm log}_{10}U$, where $dM_{\rm d}$ is the amount of dust mass heated by $U$ within a range of $\pm0.5~d{\rm log}_{10}U$. We choose $d{\rm log}_{10}U = 0.01$~dex, which is fine enough to capture the details of the distribution function, but not too fine that noise dominates the plots.

We normalize $dM_{\rm d}$ by the total dust mass, $M_{\rm Tot}$, in each galaxy, and define the mass fraction as $dm_{\rm d} = dM_{\rm d}/M_{\rm Tot}$. Then, the normalized distribution function is
\begin{equation}
\label{eq:dist_func}
\frac{dm_{\rm d}}{d{\rm log}_{10}U}~(U) = \frac{1}{M_{\rm Tot}} \ \frac{dM_d}{d{\rm log}_{10}U}~(U).
\end{equation}
We interpret this normalized distribution function, $dm_{\rm d} / d{\rm log}_{10}U$, as the fraction of the total dust mass per dex of $U$, calculated in a bin with width of $d{\rm log}_{10}U$ and centered at $U$. Less formally, this is the normalized probability density of dust mass as a function of $U$.

\textbf{Uncertainty:} To assess the uncertainty in the measured distributions, we calculate the scatter in the distribution function across a set of 100 realizations of the fitting results \citep[following][]{gordon14,chiang18}. We indicate this uncertainty by a shaded area in Figures~\ref{fig:dist_13pc} and~\ref{fig:dist_167pc}, which shows the $16^{\rm th}{-}84^{\rm th}$ percentile across all realizations. The uncertainty appears larger at 167~pc resolution because lower number of pixels leads to more statistical noise.

To create the realizations, we use the relative likelihoods for each point in our $\Sigma_d{-}T_d$ grid space, and randomly draw values from that grid weighted by those relative likelihoods. We repeat this process 100 times for each pixel in each galaxy. This yields 100 realizations of the maps of $T_d$ and $\Sigma_d$ for each galaxy.

We construct the distribution functions as described above for each realized map. In each bin of the distribution function, we define the $\pm 1\sigma$ uncertainty as the difference between the $84^{\rm th}$ and $16^{\rm th}$ percentile value across all realizations in that bin.

\textbf{Parameters of the distribution:} For each distribution, we measure three quantities described as follows.

\begin{enumerate}
    \item $U_{\rm peak}$, defined as the value of $U$ that corresponds to the peak of the distribution of mass or luminosity. This is the most common value in the distribution function. It often, but not always lies, near the center of the distribution.
    \item $U_{\rm med}$, defined as the median value of $U$ in the distribution of mass or luminosity.
    \item $\mathcal{W}_U$, defined as the logarithmic width in dex of $U$, that covers the 16th-to-84th percentile of mass or luminosity sorted by $U$. 
\end{enumerate}

We indicate $U_{\rm peak}$ and $U_{\rm med}$ for each galaxy by a circle and a tick above the distributions in Figures~\ref{fig:dist_13pc} and~\ref{fig:dist_167pc}, and the width as the horizontal line, color coded by galaxy. We record their values in Table~\ref{tab:params}.

\begin{deluxetable*}{ c | c | c c c | c c c | c c c }
\tablewidth{0pt}
\tabletypesize{\scriptsize}
\tablecaption{Parameters of the distribution function at 13~pc and 167~pc resolutions with $\beta=1.8$ \label{tab:params}}
\tablehead{
\multirow{3}{*}{Galaxies} & \multirow{3}{*}{Resolutions} & \multicolumn{3}{c |}{} & \multicolumn{3}{c |}{} & \multicolumn{3}{c}{} \\
{} & {} & \multicolumn{3}{c |}{Radiation Field} & \multicolumn{3}{c |}{Dust Equilibrium Temperature} & \multicolumn{3}{c}{Dust Mass Surface Density} \\
{} & {} & log$_{10}U_{\rm peak}$ & log$_{10}U_{\rm med}$ & Width [dex] & $T_{\rm peak}$ [K] & $T_{\rm med}$ [K] & Width [K] & log$_{10}\Sigma_{\rm peak}$ & log$_{10}\Sigma_{\rm med}$ & Width [dex]
}
\startdata
\multicolumn{10}{c}{Distribution of Mass} \\
\hline
%SMC & \multirow{2}{*}{13~pc} & $0.03$ & $-0.00$ & $0.71$ & $18.18$ & $17.97$ & $5.05$ & $-1.36$ & $-1.34$ & $0.71$ \\
%LMC & & $0.27$ & $0.23$ & $0.60$ & $20.00$ & $19.72$ & $4.69$ & $-1.27$ & $-1.25$ & $0.79$ \\
SMC & \multirow{2}{*}{13~pc} & $0.03$ & $0.01$ & $0.66$ & $18.18$ & $18.08$ & $4.74$ & $-1.36$ & $-1.33$ & $0.67$ \\
LMC & & $0.27$ & $0.23$ & $0.59$ & $20.00$ & $19.73$ & $4.67$ & $-1.27$ & $-1.24$ & $0.78$ \\
\hline
%SMC & \multirow{4}{*}{167~pc} & $0.15$ & $0.13$ & $0.56$ & $19.07$ & $18.97$ & $4.13$ & $-1.74$ & $-1.53$ & $0.70$ \\
%LMC & & $0.30$ & $0.28$ & $0.49$ & $20.24$ & $20.13$ & $3.95$ & $-1.33$ & $-1.34$ & $0.69$ \\
%M31 & & $-0.01$ & $-0.16$ & $0.62$ & $17.89$ & $16.91$ & $4.07$ & $-1.69$ & $-1.66$ & $0.59$ \\
%M33 & & $-0.43$ & $-0.27$ & $0.99$ & $15.15$ & $16.17$ & $6.50$ & $-1.35$ & $-1.38$ & $0.50$ \\
SMC & \multirow{4}{*}{167~pc} & $0.16$ & $0.15$ & $0.43$ & $19.14$ & $19.10$ & $3.28$ & $-1.74$ & $-1.49$ & $0.66$ \\
LMC & & $0.30$ & $0.28$ & $0.48$ & $20.24$ & $20.15$ & $3.92$ & $-1.33$ & $-1.34$ & $0.68$ \\
M31 & & $-0.01$ & $-0.14$ & $0.60$ & $17.89$ & $17.00$ & $3.97$ & $-1.69$ & $-1.65$ & $0.58$ \\
M33 & & $-0.52$ & $-0.33$ & $1.04$ & $14.61$ & $15.76$ & $6.72$ & $-1.31$ & $-1.36$ & $0.48$ \\
\hline
\multicolumn{10}{c}{Distribution of Luminosity} \\
\hline
%SMC & \multirow{2}{*}{13~pc} & $0.18$ & $0.27$ & $0.71$ & $19.30$ & $20.01$ & $5.74$ & $-1.15$ & $-1.28$ & $0.97$ \\
%LMC & & $0.34$ & $0.47$ & $0.74$ & $20.56$ & $21.65$ & $6.52$ & $-1.11$ & $-1.08$ & $0.96$ \\
SMC & \multirow{2}{*}{13~pc} & $0.18$ & $0.26$ & $0.68$ & $19.30$ & $19.94$ & $5.45$ & $-1.15$ & $-1.26$ & $0.90$ \\
LMC & & $0.34$ & $0.47$ & $0.74$ & $20.56$ & $21.66$ & $6.52$ & $-1.11$ & $-1.07$ & $0.95$ \\
\hline
%SMC & \multirow{4}{*}{167~pc} & $0.18$ & $0.23$ & $0.38$ & $19.30$ & $19.76$ & $3.02$ & $-1.35$ & $-1.40$ & $0.72$ \\
%LMC & & $0.35$ & $0.44$ & $0.60$ & $20.64$ & $21.41$ & $5.24$ & $-1.28$ & $-1.21$ & $0.74$ \\
%M31 & & $0.04$ & $0.03$ & $0.58$ & $18.25$ & $18.19$ & $4.17$ & $-1.43$ & $-1.63$ & $0.66$ \\
%M33 & & $0.45$ & $0.26$ & $0.79$ & $21.48$ & $19.98$ & $6.05$ & $-1.19$ & $-1.26$ & $0.49$ \\
SMC & \multirow{4}{*}{167~pc} & $0.18$ & $0.23$ & $0.37$ & $19.30$ & $19.76$ & $2.93$ & $-1.35$ & $-1.38$ & $0.69$ \\
LMC & & $0.35$ & $0.44$ & $0.60$ & $20.64$ & $21.42$ & $5.23$ & $-1.28$ & $-1.21$ & $0.74$ \\
M31 & & $0.04$ & $0.03$ & $0.57$ & $18.25$ & $18.21$ & $4.12$ & $-1.43$ & $-1.63$ & $0.67$ \\
M33 & & $0.44$ & $0.27$ & $0.82$ & $21.39$ & $20.00$ & $6.23$ & $-1.19$ & $-1.24$ & $0.45$ \\
\enddata
\tablecomments{We use Equation~\ref{eq:u2t} to convert \temp\ to $U$ (assuming $\beta = 1.8$). See text for the definition of each parameter. $\Sigma_d$ is in \sunit.}
\end{deluxetable*}

\subsubsection{Distribution of Dust Mass as a function of Radiation Field}

In practice, we calculate $dm_d/d{\rm log}_{10}U$ by summing the dust mass within each 0.01~dex-wide bin of log$_{10}U$. Then, we divide the mass in each bin by the total dust mass in the galaxy and by the 0.01~dex of bin width. For bins of $U$ that have no mass in it, we assign the value of $dm_d/d{\rm log}_{10}U$ in those bins through linear interpolation. Before plotting, we apply Gaussian smoothing with width of five bins. We plot these distributions in the {\it top left panels} of Figures~\ref{fig:dist_13pc} and~\ref{fig:dist_167pc}.

Because our estimate of $U$ depends directly on our best-fit of dust temperature, \temp, via Equation~\ref{eq:u2t}, our measured distribution function can also be converted to the distribution of mass as a function of \temp\ via
\begin{equation}
\frac{dm_{\rm d}}{d{\rm log}_{10} T_{\rm d}} = (4+\beta) \frac{dm_{\rm d}}{d{\rm log}_{10}~U}.
\end{equation}
In the {\it top left panels} of Figures~\ref{fig:dist_13pc} and~\ref{fig:dist_167pc}, we show these distribution functions in SMC and LMC at a common 13~pc resolution and in four targets at a common 167~pc resolution.

At 167~pc resolution, M31 has log$_{10}U_{\rm peak}\approx-0.01$~dex (corresponds to $T_{\rm peak} \approx 18$~K). This is very close to the Solar neighborhood value of log$_{10}U_\odot=0$~dex. But, as the top left plot in Figure \ref{fig:dist_167pc} shows, $U_{\rm peak}$ in M31 lies at the upper end of a wide distribution (i.e. $U_{\rm peak} > U_{\rm median} \approx -0.14$). Therefore, M31 includes a large amount of mass illuminated by radiation fields below the Solar Neighborhood value. This agrees with the well-known result that the integrated SED in M31 indicates cooler dust temperatures than those in the other Local Group galaxies \citep[e.g.,][]{haas98,groves12,smith12,draine14}. As Figure \ref{fig:maps_temp} shows, much of this cooler material lies in the outer part of the galaxy.

At $167$~pc resolution, M33 shows the lowest log$_{10}U_{\rm peak}$ in our sample\footnote{Keep in mind that we use a constant value of the dust emissivity index, $\beta=1.8$, for all galaxies. Since \temp\ and $\beta$ are anti-correlated, a lower value of $\beta$ in M33 \citep[$\beta=1.5$;][]{xilouris12} would lead to higher $U_{\rm peak}$. Conversely, a higher value of $\beta$ in M31 \citep[$\beta \gtrsim 1.9$;][]{smith12} would lead to a lower $U_{\rm peak}$.} ($-0.52$~dex), but this peak lies towards the low end of a wide distribution (i.e. $U_{\rm peak} <U_{\rm median}\approx-0.33$). M33 has both more mass at high~$U$ and more mass at low~$U$ compared to M31 (shown as their horizontal lines), consistent with strong dust temperature gradient \citep[e.g.,][]{xilouris12} and extended gas disk in that galaxy \citep[e.g,][]{koch18}. Again, Figure \ref{fig:maps_temp} shows that the temperature and luminosity have a stronger concentration towards the inner part of the galaxy than the dust mass.

At $167$~pc resolution, both Magellanic Clouds show narrower distributions and higher $U_{\rm peak}$ than M31 and M33. The LMC has the highest log$_{10}U_{\rm peak}$ (0.30~dex) in our sample, while the SMC has log$_{10}U_{\rm peak}$ of 0.16~dex. These values are close to their log$_{10}U_{\rm med}$ values (0.28~dex for LMC and 0.15~dex for SMC). Hence, both galaxies have most of their mass above the Solar neighborhood value ($U_\odot=1$). The LMC includes a significant ``tail'' toward high $U$, indicating a substantial fraction of its dust mass in high radiation field regions like that around 30 Doradus (see \S\ref{sec:reg} and the maps in Figure \ref{fig:maps_temp}). Compared to the two spirals (M31 and M33), the maps of the Magellanic Clouds show less mass in an extended, cool disk (Figure \ref{fig:maps_temp}).
% A similar, but more subtle tail towards high $U$ can also be seen in the SMC distribution by mass.

\pagebreak
\subsubsection{Distribution of Luminosity as a function of Radiation Field}

We calculate the distribution of equilibrium IR luminosity as a function of $U$, $d\mathcal{J}_{\rm IR} / d{\rm log}_{10} U~(U)$, analogous to the calculation of mass distribution as a function of $U$. We normalize the distribution by the integrated luminosity for the whole galaxy, $\mathcal{J}_{\rm eq,Tot}$. The resulting distribution shows the fraction of the equilibrium luminosity per dex of log$_{10}U$. We plot these distributions in the {\it top right panels} of Figures~\ref{fig:dist_13pc} and~\ref{fig:dist_167pc}.

As one might expect, given the dependence of $\mathcal{J_{\rm IR}}$ on $U \propto T_{\rm d}^{5.8}$ (Equation~\ref{eq:tir}), higher fraction of the luminosity distributions shift towards higher $U$ compared to the mass distributions. This strongly affects $U_{\rm peak}$ in M33, which has the lowest $U_{\rm peak}$ and $U_{\rm med}$ by mass but has the highest $U_{\rm peak}$ by luminosity (0.44~dex). The other galaxies show modest shifts in $U_{\rm peak}$ and $U_{\rm med}$, but the shape of their luminosity distributions around the peak change significantly (compared to that in the mass distribution). In both Magellanic Clouds, the luminosity distribution shows more prominent ``bumpy'' features at high $U$ than in the mass distribution. These reflect large contributions to the luminosity, but smaller contributions to the mass, from hot star forming regions (see \S\ref{sec:reg}).

At 167~pc resolution, M31 has the lowest log$_{10}U_{\rm peak}$ (0.04~dex) and lowest log$_{10}U_{\rm med}$ (0.03~dex), again, close to the Solar neighborhood value. Compared to the mass, less luminosity comes from the low temperature, extended part of the galaxy. This is evident where log$_{10}U_{\rm med}$ for luminosity is $0.03$, but only $-0.14$ for the mass distribution (i.e. lower value of $U_{\rm med}$ means low-$U$ regions contribute more toward the overall distribution). The prominent low-$U$ feature in the mass distribution appears dramatically suppressed in the luminosity distribution. Meanwhile, the central part of the distribution ($U \sim 1$), that maps to the star-forming 10~kpc ring in M31 (Figure~\ref{fig:maps_temp}), increases in prominence. The high-$U$ bulge region contributes much more luminosity than mass, but still represents only a small fraction of the galaxy's total luminosity (1.6\%; see \S\ref{sec:reg} for detailed discussions).

M33 shows the most striking change between the mass and luminosity distributions. Its $U_{\rm peak}$ shifts from the lowest by mass ($-0.52$~dex) to the highest by luminosity ($0.44$~dex). Its $U_{\rm med}$ also increases by 0.60~dex. This reflects the strong dust temperature gradient \citep{xilouris12}, which creates a wide range of $U$ and \temp, along with the extended gas and dust surface density distributions \citep{koch18}. As the maps in Figure~\ref{fig:maps_temp} show, the mass in M33 extends out to large radii, while the luminosity appears more centrally concentrated, coincides with active star formation (\S\ref{sec:reg}).

We note that in SMC and LMC, there is a fraction of IR luminosity that comes from dust with $T_d > 30$~K in the star forming regions (high end tail in the top right panel of Figure~\ref{fig:dist_13pc}). Even though warm dust component can contribute towards these hot regions, it only contributes a negligible fraction in the mass distribution and lies outside 68\% of the mass distribution (top left panel of Figure~\ref{fig:dist_13pc}). Hence, the accuracy of our modeling for this warm dust component does not affect the main results of this paper.

\subsubsection{The Width of the Distributions}

At 167~pc resolution in the Magellanic Clouds and M31, $68\%$ of both the mass and IR luminosity are concentrated within a $\lesssim 0.6$~dex range ($\pm 0.3$~dex, or about a factor of two). In the SMC, the distribution of luminosity appears even narrower, with $68\%$ coming from a $0.37{-}0.43$~dex wide range.

M33 yields the widest distributions of both mass (1.04~dex) and luminosity (0.82~dex) as a function of $U$. As discussed above, these wide distributions reflect the structure of M33 that is visible in Figure~\ref{fig:maps_temp}. High $\Sigma_{\rm d}$ extend into the outer disk, while \temp\ tends to drop with radius. Meanwhile, in the inner part of the galaxy, multiple knots of star formation activity exhibit both high \temp\ and $\Sigma_d$.

At $13$~pc resolution, available only for the Magellanic Clouds (Figure \ref{fig:dist_13pc}), the distributions appear wider than at $167$~pc resolution. These distributions at 13~pc resolution show more material at low radiation fields. This could be expected if the convolution to coarser resolution blurs together warm and cool dust. At their coarser resolution, the light from nearby warmer dust would make it harder to isolate the cool component, leading to the lower amount of low-$U$, low-\temp\ materials at $167$~pc resolution. This idea of relatively ``hidden'' cool dust, masked by the presence of more luminous material nearby, has appeared many times in the literature \citep[e.g.,][]{galliano05}. We explore this effect quantitatively in $\S$\ref{sec:res}.

\subsection{Distributions of Dust Mass and Luminosity as a Function of Dust Mass Surface Density}
\label{sec:dis_mass_by_mass}

We also build the distributions of mass and luminosity as a function of the dust mass surface density, $\Sigma_d$. We plot these $dm_d/d{\rm log}_{10}\Sigma_d$ and $d\mathcal{J}_{\rm IR}/d{\rm log}_{10}\Sigma_d$ in the lower panels of Figures \ref{fig:dist_13pc} and \ref{fig:dist_167pc}. We construct these distributions in the same way as those treating $U$ as the independent variable (in $\S$\ref{sec:dis_mass_by_u}), but here, we bin it by log$_{10}\Sigma_d$ instead of log$_{10}U$.

As in $\S$\ref{sec:dis_mass_by_u}, we characterize the distributions with three parameters, $\Sigma_d$ at the peak of the distribution ($\Sigma_{\rm peak}$), the median of the distribution ($\Sigma_{\rm med}$), and the logarithmic width of the distribution, $\mathcal{W_\Sigma}$, defined as the 16th-to-84th percentile range of the distribution. We show these $\Sigma_{\rm peak}$ and $\Sigma_{\rm med}$ as the dots and ticks above the distributions in Figures~\ref{fig:dist_13pc} and~\ref{fig:dist_167pc}.

\subsubsection{Mass Distribution as a function of Dust Mass Surface Density}

The {\it bottom left panel} in Figure \ref{fig:dist_167pc} shows the dust mass distribution for our targets at 167~pc resolution. Dust mixed with atomic gas, molecular gas, and even ionized gas are all contribute to these histograms, weighted by the local gas-to-dust ratio. In that sense, this plot also shows the column density distribution of the whole ISM across the Local Group.

Almost all dust mass in all four targets lies in the range $\Sigma_d \sim 0.01{-}0.1$~\sunit. For a Galactic gas-to-dust ratio (GDR) of $\sim 150$, this range equates to a gas mass surface density of $\sim 1{-}15~M_\odot$~pc$^{-2}$. The metallicities of the LMC, SMC, and M33 are all lower than that in the Milky Way \citep[e.g.,][]{russell92,rosolowsky08}, so the associated range of gas mass surface densities should in fact be more like $\sim 3{-}50~M_\odot$~pc$^{-2}$. This range is consistent with the fact that the ISM in all of these galaxies is known to be dominated by atomic gas across the disk, with a few dense regions (e.g., the inner part of M33 and the LMC's ridge region) locally dominated by molecular gas \citep[e.g.,][]{druard14,wong11}.

The LMC and M33 show similar mass distributions. They exhibit the highest $\Sigma_{\rm peak}$ of $\approx 0.045~M_\odot$~pc$^{-2}$ in our sample (their $\Sigma_{\rm peak} \approx \Sigma_{\rm med}$) and widths of $0.5{-}0.7$~dex. These two dwarf spirals have comparable metallicity, $Z \approx 0.5~Z_\odot$ \citep{russell92,rosolowsky08}, and the distribution appears consistent with dust being mostly mixed with high column density of {\sc Hi} in both targets.

The SMC shows low $\Sigma_{\rm peak}$ and low $\Sigma_{\rm med}$ of $0.02$ and $0.03$~\sunit, respectively. The SMC has a lower metallicity and higher GDR of $\sim 1200$, compared to those in M33 and the LMC \citep[$\approx 380$;][]{duval14}. The lower $\Sigma_{\rm peak}$ in the SMC could thus be expected if the gas mass surface density in SMC were comparable to that in LMC. In reality, the SMC's elongated structure along the line of sight \citep{stanimirovic99,staveley-smith03} leads to higher {\sc Hi} column densities than one finds in the other Local Group targets. These effects combine to produce the multi-components distribution (bumps) that mostly overlaps the other targets in the bottom left panel of Figure \ref{fig:dist_167pc}. We emphasize that this overlap is at least partially a coincidence due to the SMC's orientation.

M31 also shows low $\Sigma_{\rm peak}$ and $\Sigma_{\rm med} \sim 0.02$~\sunit, again indicating a large amount of mass in an extended disk with relatively low column densities and a GDR that increases with radius \citep[e.g.,][]{draine14}. Note that M31's appearance in the plot depends on the substantial correction that we have applied to account for M31's high inclination \citep[$77^{\circ}.7$;][]{corbelli10}.

\subsubsection{Luminosity Distribution as a function of Dust Mass Surface Density}

The distribution of luminosity, $\mathcal{J_{\rm IR}}$, shifts toward higher $\Sigma_d$ compared to the distribution of mass. The contribution of mass from low $\Sigma_d$ is suppressed in the luminosity distribution, changing the shape of distribution in M31 and M33 from symmetric, flat/roundish top, to have narrow peak with tail at low end (positively skewed). The main peak in the luminosity distribution for M31 mostly comes from the star-forming ring and less from the low $T_d$ in the outer disk. The tail towards low $\Sigma_d$ shows the influence of the hot bulge.  

In the LMC, the distinct components (bumps) in the mass distribution become more pronounce in the luminosity distribution, reflecting the fact that warm dust in high-$\Sigma_d$ star-forming regions contribute a lot to the IR luminosity. In the SMC, the distribution changes from negatively skewed in mass to a more symmetric in luminosity, because the high mass surface density regions also tend to have warmer temperature that emits a large fraction of the luminosity.

Because IR luminosity proportional to $\Sigma_d T_d^{5.8}$, the difference between the mass and luminosity distributions implies that temperature correlates with $\Sigma_d$, i.e. part of the luminosity distribution from low $\Sigma_d$ is suppressed (because they also have lower \temp), while luminosity distribution from high $\Sigma_d$ is enhanced (because they also have higher \temp). In $\S$\ref{sec:pix} we show that this appears to be the general case in high signal-to-noise regions, though with some notable exceptions.

\subsubsection{Widths Comparisons}

At 167~pc resolution, the SMC and LMC have similar width in the distributions of mass and luminosity ($\sim 0.7$~dex), and only slightly larger than the width for M31 and M33 ($\sim0.5{-}0.6$~dex). This means the SMC and LMC have the largest variations in $\Sigma_d$ where 68\% of the mass is distributed. As in the mass distribution, the width of luminosity distribution at 167~pc is smaller than that at 13~pc resolution (Table~\ref{tab:params}) because of convolution between regions with low and high mass surface density.

The shapes of the mass distributions somewhat resemble the distributions of molecular gas mass as a function of $\Sigma_d$ in nearby galaxies seen at similar spatial resolution by \citet{sun18}. They also found signature of multiple components, usually related to distinct dynamical regions (bulge vs. disk). The distributions that we find show more width and  more components, perhaps reflecting the wider mixture of environments (e.g. spiral arms and star-forming regions) that is incorporated into our distribution.

In \S\ref{sec:reg}, we will see that the distributions for any isolated region may appear roughly lognormal in shape. This would agree with the observation that the column density distribution of diffuse neutral gas density in the Milky Way and other Local Group galaxies has a lognormal shape \citep{berkhuijsen08,berkhuijsen15,corbelli18}. The same shape also emerges from simulations \citep{wada07}. Making a more rigorous measurement of the shape of the distribution after controlling for regional variations is a subject for future work.

\subsubsection{Implications}

These mass and luminosity distributions as a function of $\Sigma_d$ have two implications: low infrared optical depth and no evidence for opaque {\sc Hi}, as described below.

{\bf Low Infrared Optical Depth:} The dust mass surface densities in our targets almost never exceed $\Sigma_{\rm d} \sim 1$~M$_\odot$~pc$^{-2}$. Given our adopted value of $\kappa$ at 160~\micron, this implies that at our resolutions, the optical depth throughout our targets at 100 and 160~\micron\ will be $\lesssim 1 \times 10^{-2}$ and $5 \times 10^{-3}$, respectively. This validates our assumption of optically thin dust that is used in the model (i.e. approximating Equation~\ref{eq:rad_trans} by Equation~\ref{eq:opt_thin}).

The opacity at $100{-}160$~\micron\ appears small at both $13$~pc (for the Magellanic Clouds) and $167$~pc (for all targets). The importance of pressure from reprocessed IR emission scales as the infrared opacity, $\tau_{IR}$ \citep[e.g., see][]{thompson05}, and should exceed the UV-driven radiation pressure when $\tau_{\rm IR}>1$. Thus, we expect radiation pressure from \textit{long wavelength} infrared photons will contribute an {\it insignificant} amount of feedback over essentially all of the area in these targets. 

The highest $\Sigma_d$ that we observe in the LMC (Figure~\ref{fig:dist_13pc}) imply significant opacity at shorter IR wavelengths (e.g., $\lambda \sim 20$~\micron). We do not consider emission at these wavelength. However, given the surface densities that we observe and the fact that star forming regions often produce significant emission at these mid-IR wavelengths, our observations are consistent with a significant effect from reprocessed infrared photons in the hottest, highest density parts of our targets \citep{lopez11,lopez14}. To see this, note that at $13$~pc resolution, the LMC by luminosity includes a small, but noticeable, contribution at $\Sigma_d \sim 1{-}10$~M$_\odot$~pc$^{-2}$ (bottom right panel of Figure \ref{fig:dist_13pc}). This $\Sigma_d \sim 1$~M$_\odot$~pc$^{-2}$ implies $\tau_{\rm 160 \micron}$ of only $\sim 4 \times 10^{-3}$. If we extrapolate to $\sim 20$ \micron, though, $\tau_{\rm 20 \micron} \sim 0.15{-}0.3$ (the extrapolation using $\beta \sim 1.8{-}2$ is quite approximate but not unreasonable, see \citealt{draine11}). Then, even at the extreme tail of the LMC distribution, we should see a non-trivial, and perhaps even dominant, contribution to the radiation pressure from IR-reprocessed photons.

{\bf No Evidence for Hidden Opaque {\sc Hi}:} The modest $\Sigma_{\rm d}$ values that we observe and the smooth appearance of our $\Sigma_{\rm d}$ maps provide evidence {\it against} the existence of the patchy, high column density, opaque {\sc Hi} clouds posited by \citet{braun09} and \citet{braun12}. \citeauthor{braun12} presented opacity-corrected {\sc Hi} column density maps in three of our targets; the LMC, M31, and M33. The 21-cm maps that were used to create these {\sc Hi} maps have comparable angular resolution to our $\Sigma_{\rm d}$ maps. In \citeauthor{braun12}'s maps, the high gas column density features ($N_{\rm H} \sim 10^{23}$~cm$^{-2}$) appear as small patches sprinkled throughout the maps. 

In contrast, our $\Sigma_{\rm d}$ maps do not have a similar patchy appearance, nor do the values of $\Sigma_{\rm d}$ that we find support the presence of gas with column densities approaching $10^{23}$~cm$^{-2}$, scattered through the Local Group at this resolution. For the LMC, M33, and M31, we expect the gas to dust ratios of $\sim 100{-}500$ \citep{leroy11,duval14}. In those cases, the $\Sigma_{\rm d}$ that we find corresponds to column densities of $< 5 \times 10^{22}$ cm$^{-2}$. In order for the opaque {\sc Hi} features posited by \citet{braun12} to exist but not appear in our $\Sigma_{\rm d}$ maps, they would need to be unusually dust poor (i.e. gas to dust ratio of $> 1,000$). To our knowledge, there is no plausible mechanism to produce dense, opaque atomic gas clouds that are preferentially depleted in dust.

Our results do not rule out 21-cm line opacity playing an important role in galaxies, even in these galaxies. More modest effects that operate smoothly across the maps would be hard to distinguish from variations of the gas-to-dust ratio or the presence of CO-dark molecular gas. This issue certainly remains an open topic. Our maps simply do not support with specific structure found by the \citeauthor{braun12}'s analysis.

\subsection{Regional Comparisons}
\label{sec:reg}

As we see in Figures~\ref{fig:dist_13pc} and~\ref{fig:dist_167pc}, there are clear differences in the shape of distribution between mass (left panels) and luminosity (right panels). There are also multiple components in some of the distributions. Comparing the distributions to the maps in Figure \ref{fig:maps_temp} suggests that some of these features arise from prominent star-forming regions. Features associated with ongoing massive star formation appear particularly prominent in the luminosity maps of all four targets.

To make a quantitative connection between the structure in the maps and the distribution functions, we divide each galaxy into ``star-forming'' and ``non star-forming'' regions and build separate distribution functions for each. We normalize these mass and luminosity distributions by the total mass and total luminosity for the whole galaxy (as in \S \ref{sec:dis_mass_by_u} and \S \ref{sec:dis_mass_by_mass}). For M31, we also build separate distributions for the hot, low $\Sigma_d$ central region (inside 1~kpc from the nucleus).

\begin{figure*}
\centering
\includegraphics[width=\textwidth]{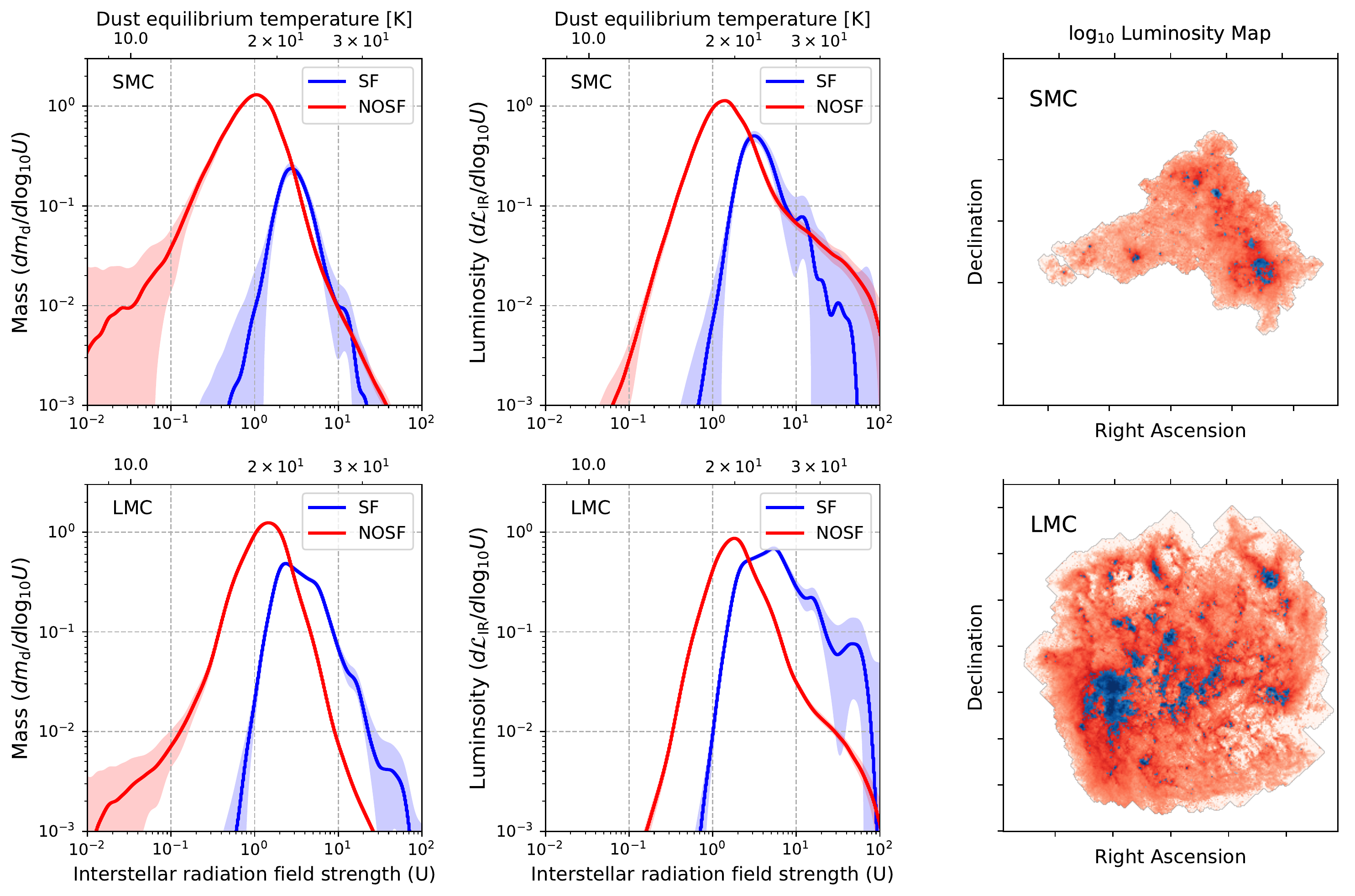}
\caption{{\bf Comparisons of distribution function for SF regions only (blue) and the rest of the galaxy (red) in the Magellanic Clouds at 13~pc resolution.} Left panels: mass distributions, middle panels: luminosity distribution, and right panels: maps of the SF regions (blue) defined as pixels in MIPS 24\micron\ with fluxes $> 1$ MJy sr$^{-1}$. The difference between blue and red distributions occurs in the high $U$ regime, which is expected because the star-forming regions have warmer dust.}
\label{fig:no_pdr_13}
\end{figure*}

\begin{figure*}
\centering
\includegraphics[width=\textwidth]{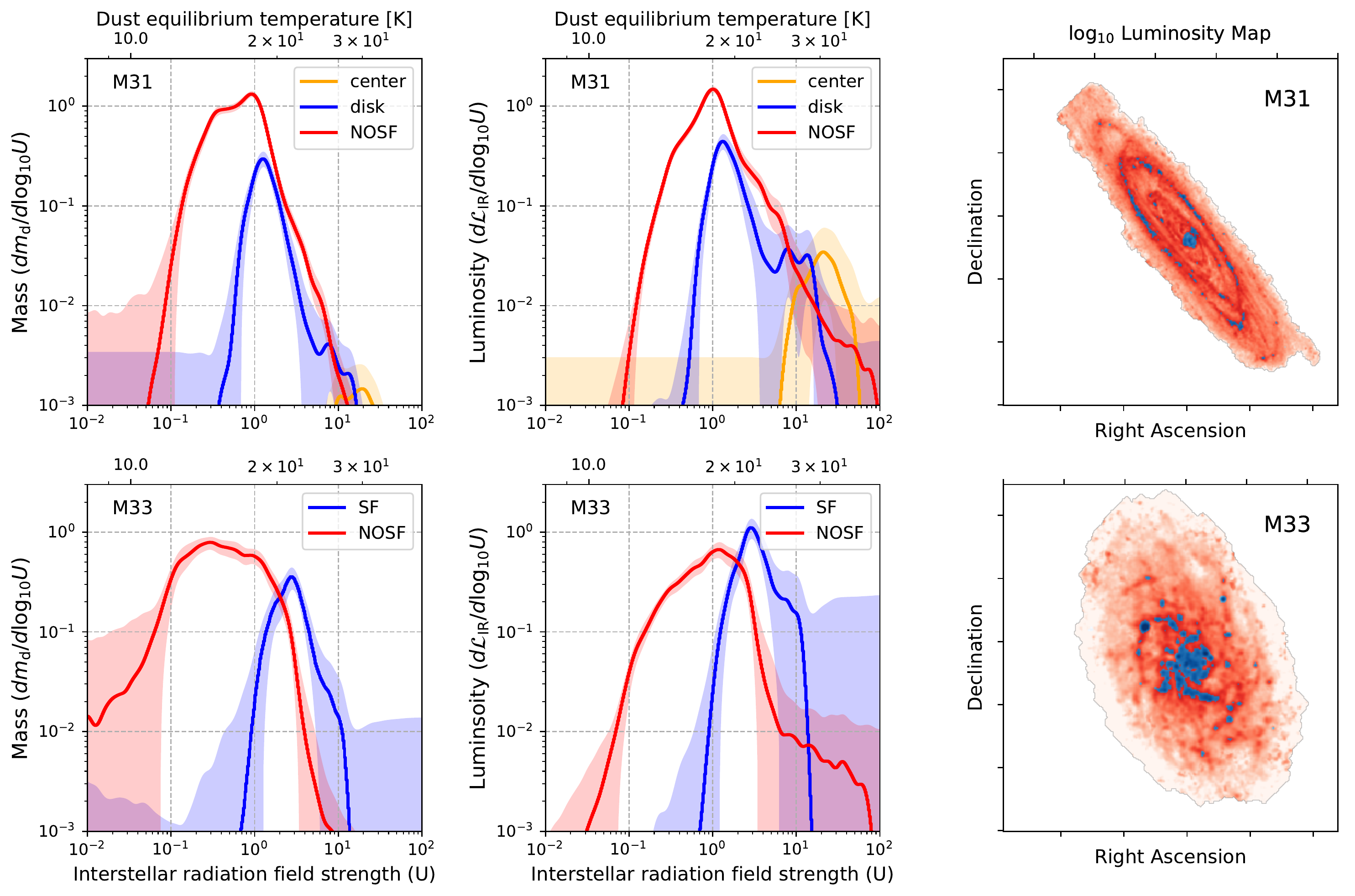}
\caption{{\bf Comparisons of distribution function for SF regions only (blue) and the rest of the galaxy (red) in M31 and M33 at 167~pc resolution.} Left panels: mass distributions, middle panels: luminosity distribution, and right panels: maps of the SF regions (blue) defined as pixels in MIPS 24\micron\ with fluxes $> 1$ MJy sr$^{-1}$.}
\label{fig:no_pdr_167}
\end{figure*}

{\bf Selection of Regions:} We define ``star-forming'' (SF) regions as areas with {\it Spitzer} MIPS \citep{werner04,rieke04} 24~\micron\ intensity $>1$~MJy~sr$^{-1}$ (shown as the blue regions in the right panels of Figures~\ref{fig:no_pdr_13} and~\ref{fig:no_pdr_167}). This intensity is equivalent to $\approx7\times10^{-4}~M_\odot$~yr$^{-1}$~kpc$^{-2}$ \citep[using the calibration from][]{calzetti07}. We choose this particular value because it selects the bright regions in the star-forming ring of M31 \citep{taba10} and star-forming complexes in the Magellanic clouds and M33 \citep{relano09}.

This cutoff also selects the bulge of M31, which shows hot dust (in Figure \ref{fig:maps_temp}) that \citet{groves12} and \citet{smith12} argued to be heated by the older stellar population. Therefore, we build separate distributions for this central region \citep[defined as the area within 1~kpc from the nucleus,  based on a position angle of $37^{\circ}.7$ from][]{corbelli10}.

{\bf Resulting Distributions:} We show the mass and luminosity distribution functions for star-forming regions (blue) and the rest of area in the galaxy (red) in Figures~\ref{fig:no_pdr_13} and \ref{fig:no_pdr_167} (with a log-scale in the vertical axis).

In all four galaxies, both the mass and the luminosity distributions from star-forming regions appear distinct compared to the rest of area in the galaxy. The star forming regions tend to show narrower distributions with a few prominent peaks. As expected, in all four galaxies, the star-forming regions tend to have warmer dust temperature than the rest of area in the galaxy.

Those figures also show that multi-component distributions in Figures~\ref{fig:dist_13pc} and~\ref{fig:dist_167pc} are a result from superimposing physically distinct regions. Star-forming regions, which also tend to be the regions rich in molecular gas, contribute many of the high-$\Sigma_d$ and high-$U$ features in the full-galaxy distributions. Because the star forming regions have both high $T_d$ and high $\Sigma_d$, they appear even more prominent in the luminosity distribution.

To be concrete, for our adopted definition, the fraction of dust mass that resides in the star forming regions is 10.0\% in the SMC, 28.4\% in the LMC, 10.2\% in M31, and 13.9\% in M33. The fraction of luminosity from the star-forming regions is higher, i.e. 24.3\% in SMC, 51.4\% in LMC, 19.4\% in M31, and 44.8\% in M33.

The bulge in M31 has higher dust temperature than the star-forming ring or the rest of the galaxy ($\approx 30$~K vs. 17~K), but it only captures very small fraction of the total dust mass ($\approx 0.08\%$). However, this central region (orange histogram in Figure~\ref{fig:no_pdr_167}) becomes more prominent in the luminosity distribution, with luminosity fraction of $1.6\%$. This more prominent contribution in the luminosity distribution, in general, is also true for star-forming regions.

\pagebreak
\section{Resolved Correlations Among Dust Parameters} \label{sec:corr}

\subsection{Correlation between Dust Temperature and Mass Surface Density}
\label{sec:pix}

\begin{figure*}
\centering
\epsscale{0.85}
\plotone{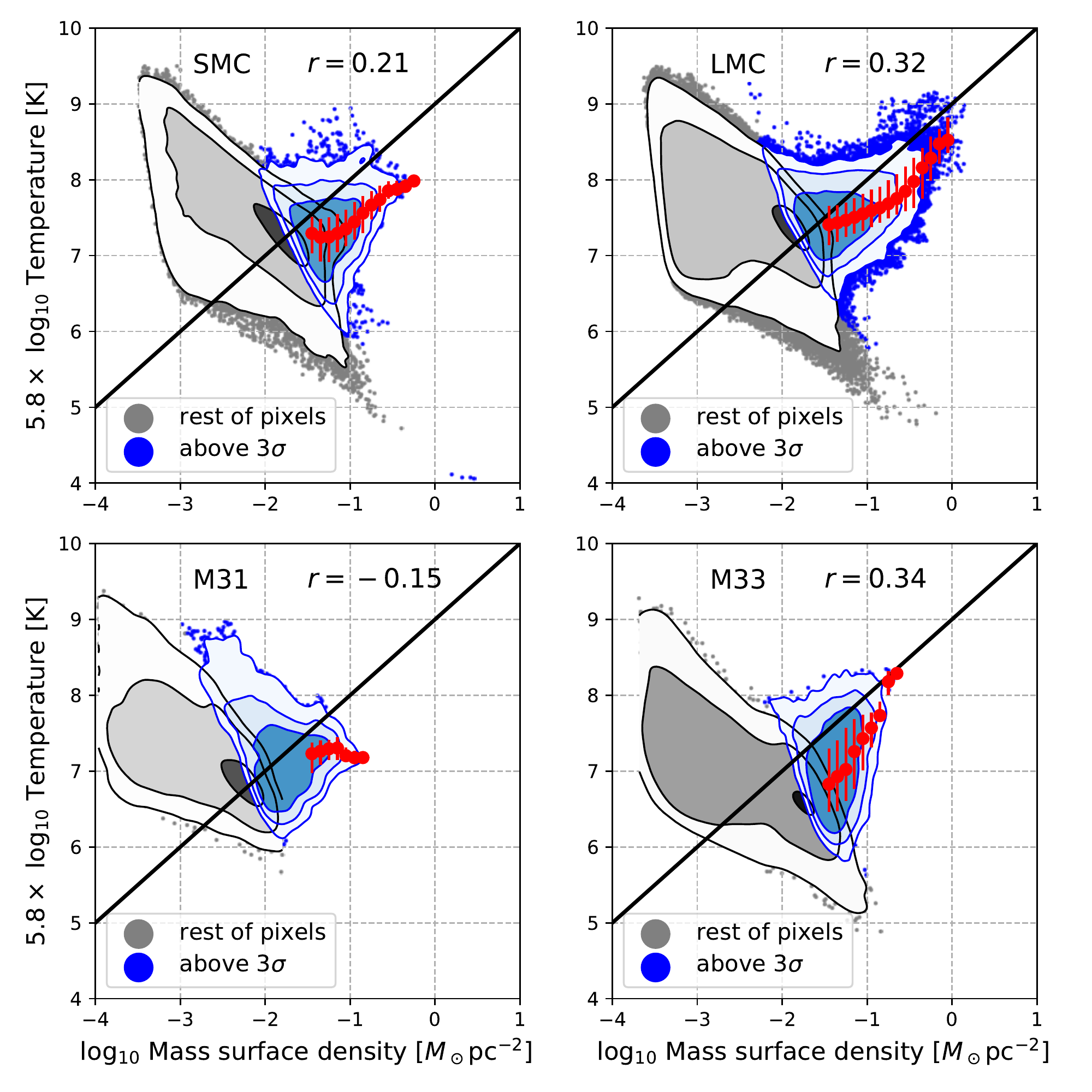}
\caption{{\bf Correlation between dust equilibrium temperature and dust mass surface density for each galaxy.} The blue points are pixels above three times the uncertainty, while the rest of pixels are shown as grey points. The grids for fitting are in the range of $5-50$~K for temperature and $10^{-4}-10$ \sunit\ for dust mass surface density. The grey contours mark the probability density of data points of 0.01, 0.1, and 0.5 points per grid, while the blue contours mark the probability density of data points of 0.01, 0.1, and 1.0 points per grid. The red points show the median temperature in 0.1~dex bin in log$_{10}\Sigma_d$, where log$_{10}\Sigma_d > -1.5$. The solid black line show a relation of 5.8~log$_{10}T_d \approx 9+$log$_{10}\Sigma_d$. The \citet{spearman04} rank correlation coefficients, $r$, for all blue points are indicated. In the SMC, LMC, and M33, the temperature and mass for blue points are correlated, where high mass surface density tends to be located in the warm star-forming regions. However, in M31, dust temperature is uncorrelated with the dust mass surface density.}
\label{fig:mass_temp}
\end{figure*}

Both the maps (Figure \ref{fig:maps_temp}) and the distributions (Figures \ref{fig:dist_13pc} to \ref{fig:no_pdr_167}) suggest a correlation between $\Sigma_d$ and $T_d$, so that high mass surface density regions also tend to be hotter and more luminous. Figure~\ref{fig:mass_temp} shows this relationship directly, where we plot dust temperature against the dust mass surface density pixel-by-pixel at 13~pc resolution for the Magellanic Clouds and 167~pc resolution for M31 and M33.

At low signal to noise, we expect the correlated uncertainties to drive an apparent anti-correlation between the best fit $\Sigma_d$ and $T_d$. This reflects that for higher temperature, less mass will be required to produce any given intensity (Equation~\ref{eq:mbb}). To suppress this effect, we plot pixels with signal-to-noise less than $3$ in either \temp\ or $\Sigma_d$ as gray points and contours. For the rest of this section, we focus our attention on the blue points and contours, which have $S/N > 3$ in both quantities and should be less affected by noise.

Because the maps contain many individual pixels, we use the Gaussian kernel density estimate in {\tt Scipy} \citep[with bandwidth estimate following][]{scott92} to calculate the probability density of data points in $200\times200$ grids between the minimum and maximum values of those data points. The contours in Figure~\ref{fig:mass_temp} mark the probability density of 0.005, 0.01, 0.05, and 0.1 data points per grid.

For the SMC, LMC, and M33, the red points show the median temperature in 0.1~dex bin of log$_{10}\Sigma_d$, in regions where log$_{10}\Sigma_d > -1.5$. Even though the \citet{spearman04} rank correlation coefficient, $r_s$, is low for all blue points (S/N $> 3$), the weak correlation between temperature and dust mass surface density is highly significant because the $p$-value is almost zero. This means warm regions also tend to have high mass surface densities. As we saw in the last section, star forming regions in the maps tend to have both high mass surface density and high temperature. 

At some degrees, a correlation between \temp\ and $\Sigma_d$ should be expected on large scales for dust internally heated by star formation. Following \citet{schmidt59} and \citet{kennicutt98}, gas with high mass surface density forms stars more rapidly. If the light from young stars is reprocessed into equilibrium IR emission, then the emergent luminosity should track this star formation. If we approximate $\Sigma_{\rm SFR} \propto \mathcal{J_{\rm IR}} \equiv \Sigma_d T_d^{5.8}$ and $\Sigma_{\rm gas} \propto \Sigma_d$, then the correlation between $\Sigma_d$ and $T_d^{5.8}$ appears when the $\Sigma_{\rm SFR}-\Sigma_{\rm gas}$ relationship is steeper than linear. Indeed, these galaxies are all mostly {\sc Hi}-dominated, and the $\Sigma_{\rm SFR}-\Sigma_{\rm HI}$ scaling relation does have a super-linear slope \citep[e.g.,][]{bigiel08,schruba11}, but also has an important dependency on other parameters \citep{leroy08}. This one-to-one correlation between $\Sigma_d$ and $T_d^{5.8}$ is shown as the black line in Figure~\ref{fig:mass_temp} (with an arbitrary normalization).

This $\Sigma_d \propto T_d^{5.8}$ scaling would be expected for dust heated by star formation. But our sample also includes several cases where this should not be a good approximation. As mentioned several times, the bulge of M31 has hot, low surface density dust. This stands out in the bottom left panel of Figure \ref{fig:mass_temp}. This dust has been shown to be externally heated by the old stellar population \citep{groves12,draine14}. 

At $13$~pc resolution, the approximation of internal heating may break down due to time evolution of star forming regions. Following \citet{schruba10} for M33, at such high resolution the heating sources and gas peaks resolve into discrete features. This effect can also be seen in the Magellanic Clouds \citep{jameson16}. In this case, the local $\Sigma_d$ will no longer be directly correlated to the amount of heating.

\pagebreak
\subsection{Intensity at 500 \micron\ and Dust Mass Surface Density} \label{sec:500_mass}

With powerful sub-millimeter telescopes like ALMA, it has become common to estimate the dust mass from only one or a few measurements on the Rayleigh-Jeans tail of the SED. This approach assumes a dust temperature, either implicitly or explicitly. Here, we compare $\Sigma_d$ from our five bands fitting, which leverages information from the $100{-}500$~\micron\ SED to the monochromatic intensity at $500$~\micron. This offers a check on the reliability of sub-mm intensities as a dust mass tracer.

\begin{figure*}
\centering
\epsscale{0.9}
\plotone{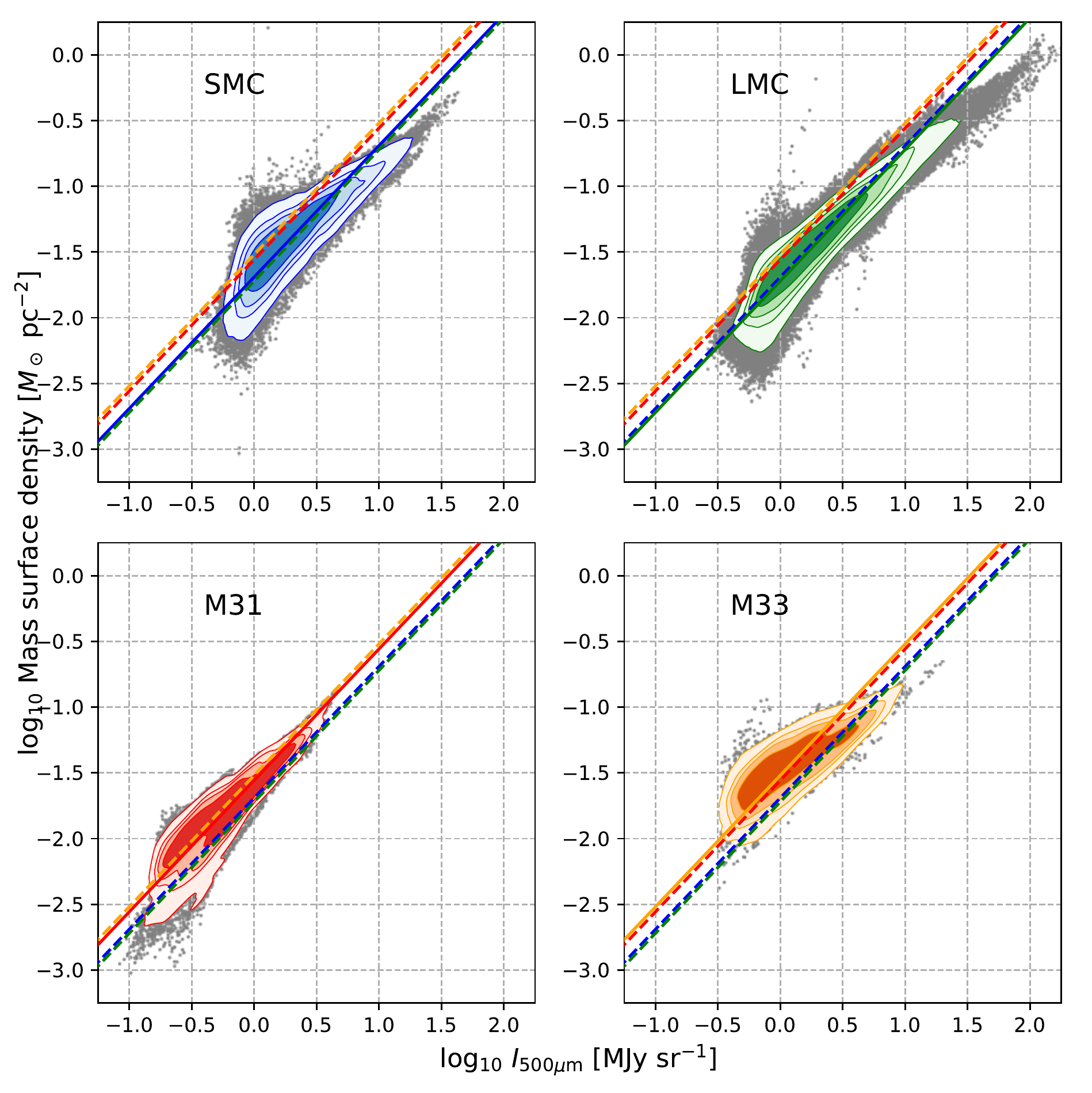}
\caption{{\bf The correlation between intensity at 500 \micron\ and the derived dust mass surface density} from the modified blackbody model with single-temperature. Contours show the density of data points, calculated using the Gaussian kernel density estimate. Different panels and colors denote different galaxies. The Magellanic Clouds is at 13~pc resolution, while M31 and M33 is at 167~pc resolution. The diagonal line has a slope equal to the median ratio of $\Sigma_d$ over $I_{500\micron}$ and intercept the median values of $\Sigma_d$ and $I_{500\micron}$. This figure shows that even a simple median ratio is a good representation of the correlation between $\Sigma_d$ and $I_{500\micron}$.}
\label{fig:corrs}
\end{figure*}

Figure~\ref{fig:corrs} shows the correlation between resolved $I_{500\micron}$ and $\Sigma_d$ over three orders of magnitude in both axes. We only select pixels above $2\sigma$ uncertainty for both $I_{500\micron}$ and $\Sigma_d$. Contours show the density of data points, generated using the Gaussian kernel density estimate with a kernel size of $\approx 0.02$~dex. In each panel, the diagonal lines show relations with power law index of 1 with normalization set by the median of $\Sigma_d/I_{500\micron}$ measured for each galaxy. The value of this median ratio is $\Sigma_d/I_{500\micron} \approx$ $1.91_{-0.43}^{+0.51} \times 10^{-2}$ for the LMC, $2.05_{-0.53}^{+0.71} \times 10^{-2}$ for the SMC, $2.84_{-0.65}^{+0.94}$ for M31, and $2.70_{-0.87}^{+1.31}$ for M33, where $\Sigma_d$ is in units of $M_\odot$~pc$^{-2}$ and $I_{500\micron}$ is in units of MJy~sr$^{-1}$. The lower and upper limits are the 16th and 84th percentiles, respectively. Among all of our targets, the variation of this ratio roughly agrees with the simple expectation based on our measured variations in $T_d$.

Both contours of M31 and M33 show a shallow slope in Figure \ref{fig:corrs} (compared to one-to-one relation). These reflect the temperature gradients discussed above \citep{xilouris12}. In these galaxies, taking $I_{500\micron}$ to linearly trace $\Sigma_d$ would give a bias so that the dust appears to be more centrally concentrated than it actually is. In other words, the temperature gradient causes the inner part of M33, and the star-forming ring and bulge of M31, to glow more brightly than the rest of the galaxy, while the cooler outer regions are fainter. This effect appears less pronounced in the Magellanic Clouds at high resolution.

Overall, this exercise confirms that $I_{500\micron}$ predicts $\Sigma_d$ within an accuracy of about $50\%$ for our sample, but with systematic biases in the two disk galaxies with temperature gradients. We emphasize that this only tests the ability to recover the \textit{dust} surface density. We have taken no account here for variations in the gas-to-dust ratio.

\subsection{Infrared Color and Temperature}

Finally, we compare the 100-to-250~\micron\ or 100-to-350~\micron\ infrared color to the $T_d$ from our five band fits. This allows us to check how well our model reproduces this IR color and to verify how well we would have predicted $T_d$ from only a single band ratio. We focus on $I_{100\micron}/I_{250\micron}$ and $I_{100\micron}/I_{350\micron}$ as colors because those wavelengths bracket the peak of the dust SED at $\sim 160$\micron.

\begin{figure*}
\centering
\epsscale{1.2}
\plotone{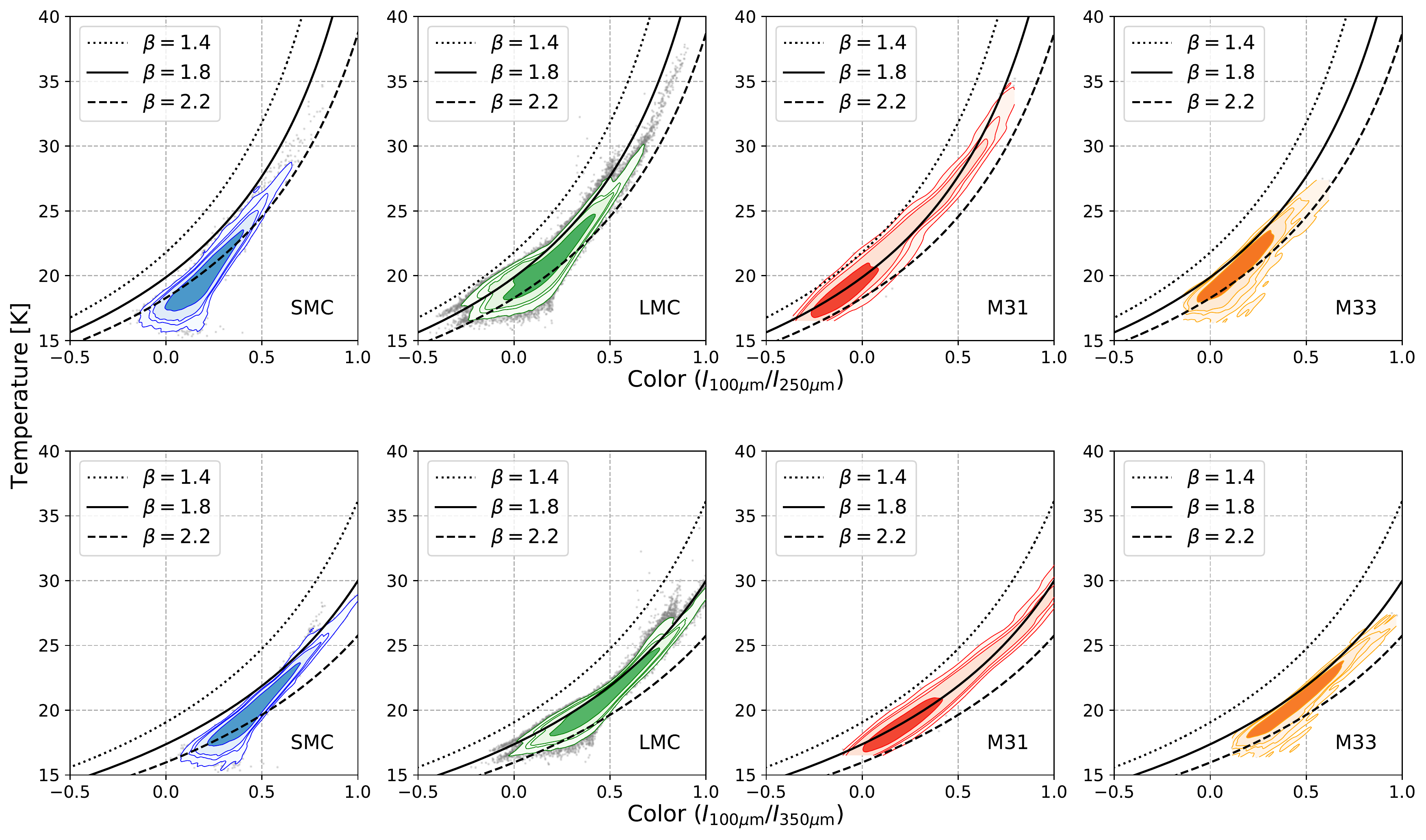}
\caption{{\bf Infrared colors against the dust equilibrium temperature.} The IR-colors in the top and bottom panels are $I_{100\micron}/I_{250\micron}$ and $I_{100\micron}/I_{350\micron}$, respectively. Contours show the density of data points estimated using the Gaussian kernel density. The black curves show the relation from the modified blackbody models with $\beta = 1.4$ (dotted), 1.8 (solid), and 2.2 (dashed). The Magellanic Clouds is at 13~pc resolution, while M31 and M33 is at 167~pc resolution. M31 data follow $\beta=1.8$ model very well but others show deviations. LMC and M33 look better to be described with $1.8<\beta<2.2$, while SMC shows variation in $\beta$, where warm regions have $\beta \sim 1.8$ and cold regions have $\beta \sim 2.2$.}
\label{fig:temp_color}
\end{figure*}

Figure~\ref{fig:temp_color} shows $T_d$ as a function of IR color for data with $S/N>10$ in both color and temperature. Contours indicate data density as in \S\ref{sec:500_mass}. We also plot the relation between temperature and colors as expected from the modified blackbody models with $\beta=$ 1.4, 1.8, and 2.2.

The data broadly follow the model, although with some notable deviations. The plots suggest that LMC and M33 may be better described by $\beta$ closer to $\sim 2$ than our adopted $\beta = 1.8$. Both the SMC and the LMC show points covering a range of $\beta$, with the regions with redder IR color are more consistent with higher $\beta$, while the rest of the pixels is in good agreement with our adopted value of $\beta$. Note that despite the shift in $\beta$, these features are as likely come from multiple dust populations unresolved by our observation (mixing of multiple populations within the beam) and the influence of out-of-equilibrium heating at short wavelengths, so that they appear as variations in dust properties (changing $\beta$). These reflect deviations from our model. The strength of this feature in the SMC appears consistent with previous observations that hot, small grains contribute heavily to the SED in this galaxy \citep[e.g.,][]{bot04}.

Lastly, we note that in the range of $30 < T_d < 40$~K, the modified blackbody models still show a curvature in the color${-}$temperature relation (black curves in Figure~\ref{fig:temp_color}). This means we can still constrain the dust temperature well within that temperature range. A vertical slope would mean the temperature loses its sensitivity to the IR color (i.e. approaching Rayleigh-Jeans tail), and this is clearly not the case in our study. Therefore, the high-temperature tail in the distributions (Figure~\ref{fig:dist_13pc}) is real.

\section{The Dependence of Dust Mass and Temperature on Physical Resolutions} \label{sec:reso}

\subsection{Dust Temperature as a Function of Physical Scale} 
\label{sec:res}

We study four of the closest star-forming galaxies at the diffraction limit of the best infrared telescope to date. This high physical resolution is not available for distant galaxies. Still, as discussed in \S\ref{sec:intro}, there remains considerable interest in using dust as a tracer of gas in more distant systems. Therefore, we explore the effects of physical resolution on the derived dust temperature and mass.

We choose the weighting methodology (described below) to create coarser resolution maps. This is different than convolving the SED first and then fitting that convolved SED. The reason behind this choice is we lose the background area gradually in the coarser resolution maps. This would make the covariance matrix calculation (Equation~\ref{eq:chi2}) becomes difficult and leading to non-robustness of the fitting results of the convolved SED (see Appendix~\ref{app: cov}).

{\bf Method:} Observing at coarse resolution tends to emphasize the sources of luminosity, rather than mass. To investigate the effects of resolution, we compare our high resolution results to what we would derive from luminosity (rather than mass) weighted averages at some coarser resolutions. We compare any results at coarser resolutions to the mass-weighted average at the finest resolution, because we consider the latter as a quantity closest to the true temperature from the majority of dust mass.

Specifically, we calculate the luminosity-weighted temperature, $T_{d,L}$,
and the luminosity-weighted mass surface density, $\Sigma_{d,L}$, at resolutions coarser than our original maps. To calculate  $T_{d,L}$, we begin with our highest resolution \temp\ map, weight that map by luminosity, convolve that map to lower resolution, and then, divide the result by the convolved luminosity map. Mathematically, this is expressed as
\begin{equation}
\label{eq:tdl}
T_{d,L} = \frac{\mathcal{J_{\rm IR}}~T_d~\ast~\mathcal{G}}{\mathcal{J_{\rm IR}}~\ast~\mathcal{G}}~,
\end{equation}
where $\mathcal{J_{\rm IR}}$ is a quantity proportional to the equilibrium dust luminosity per unit area in the highest resolution map (as defined by Equation~\ref{eq:tir}), $\ast$ denotes convolution, and $\mathcal{G}$ is the Gaussian kernel to convolve from the finest resolution to the target coarser resolution.

Similarly, we calculate the luminosity-weighted mass surface density, $\Sigma_{d,L}$, via
\begin{equation}
\label{eq:sdl}
\Sigma_{d,L} = \frac{\mathcal{J_{\rm IR}}~\Sigma_d~\ast~\mathcal{G}}{\mathcal{J_{\rm IR}}~\ast~\mathcal{G}}~.
\end{equation}

In Equations~\ref{eq:tdl} and~\ref{eq:sdl}, we calculate $T_{d,L}$ and $\Sigma_{d,L}$ for each pixel at coarser resolution. To represent the temperature for a single galaxy, we also calculate the mean value of temperature (weighted by mass) at each resolution via
\begin{equation}
\label{eq:mwtemp}
\langle T_{\rm d} \rangle_M = \frac{\sum_i~\Sigma_{d,L,i}~T_{{\rm d},L,i}}{\sum_i~\Sigma_{d,L,i}}~,
\end{equation}
Similarly, the luminosity-weighted mean temperature for a single galaxy at coarser resolution is
\begin{equation}
\label{eq:lwtemp}
\langle T_{\rm d} \rangle_L = \frac{\sum_i~\mathcal{J_{\rm IR,i}}~T_{{\rm d},L,i}}{\sum_i~\mathcal{J_{\rm IR,i}}}~.
\end{equation}
Here, the sum $\sum_i$ runs over all pixels inside the region of interest.

As we have seen in $\S$\ref{sec:reg}, weighting by luminosity leads to higher \brat\ because more weight is attached to hot, high luminosity star-forming regions. Because we use the luminosity-weighting to convolve the data to coarser resolutions, information below the resolution is irrevocably ``washed out.'' Comparing mass-weighted results at different resolutions shows how much this ``sub-resolution'' luminosity-weighting changes our inferences about the dust temperature and mass. This gives us an empirical estimate of how much the implicit luminosity-weighting in low resolution observations will wash out information on the true dust mass.

\begin{figure*}
\centering
\epsscale{0.9}
\plotone{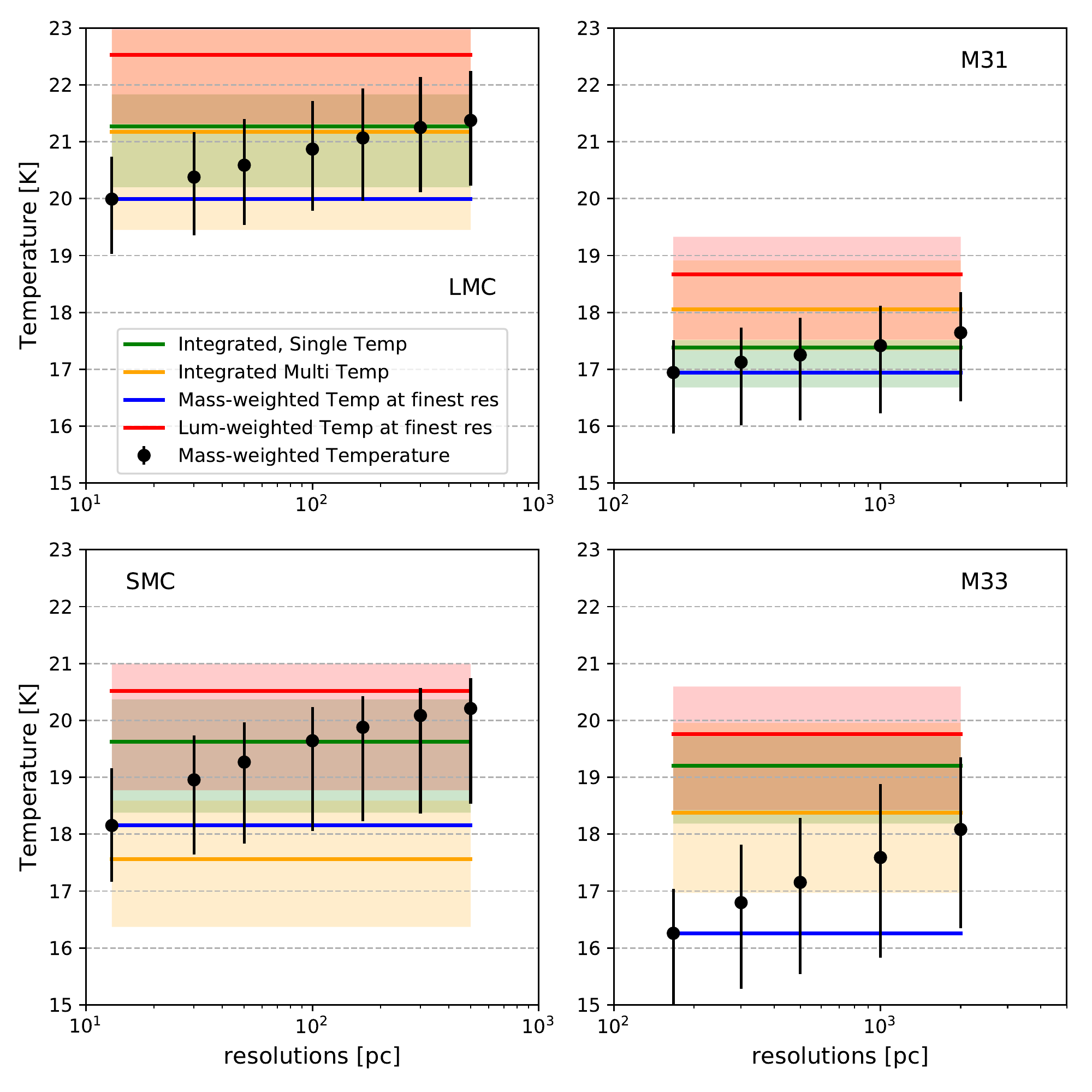}
\caption{{\bf The mass-weighted mean of dust temperature (black dots) as a function of resolution in each galaxy.} We mark the mass-weighted and the luminosity-weighted mean temperature measured at the finest resolution as the blue and red lines, respectively. The green and orange lines are the dust temperature from fitting the integrated SED with single- and multi-temperature approaches, respectively. The error bars and shaded areas are derived by taking the 16th-to-84th percentile range in the probability distribution function of the likelihood. This figure shows that the mass-weighted mean of temperature increases in coarser resolutions and approaching the luminosity-weighted average (red line) because convolution favors bright regions which are also warm.}
\label{fig:temp_res}
\end{figure*}

{\bf Results for temperature:} In Figure~\ref{fig:temp_res}, we plot \mbrat\ (black dots) for each galaxy as a function of physical resolution. We also mark \mbrat\ at the finest resolution as blue lines. This is the quantity that is closest to the true mass-weighted mean of temperature in a galaxy. Any other measurements should be compared to this line. Due to different physical sizes among galaxies, LMC and SMC are convolved to $7$ resolutions from $13$~pc to $500$~pc, while M31 and M33 are convolved to $5$ resolutions from $167$~pc to $2$~kpc.

In general, \mbrat\ rises as the resolution becomes coarser. This has the expected sense, because convolving cool, low surface density regions with warm, high surface density regions would bias \mbrat\ towards higher temperatures. We already saw an indication of this effect when comparing the distributions of mass by radiation field at $13$~pc and $167$~pc resolution for the Magellanic Clouds (Figures~\ref{fig:dist_13pc} and~\ref{fig:dist_167pc}).

This trend of \mbrat\ with spatial resolution is weakest in M31; the inferred mean temperature only changes by $< 1$~K as we blur the maps by a factor of $\sim 10$ in resolution. We identify two likely causes. First, we see a weak correlation between temperature and mass surface density in Figure~\ref{fig:mass_temp}. Second, much of the variation in temperature occurs radially on kpc scales or larger, so that blurring together the distinct physical conditions is less of an issue in M31.

The SMC and M33 show the largest increases in \mbrat\ with coarser spatial resolutions. The implied temperature in both galaxies increases by $\sim 2$~K as the resolution degrades by a factor of $\sim 10$. The LMC shows an intermediate case. We again attribute the magnitude of this resolution effect to the spatial structure and dynamic range in temperature variations in the galaxy. In general, the smaller, more star-formation dominated systems show more systematic effects, and more information loss, from blurring the data.

The red and blue horizontal lines in Figure~\ref{fig:temp_res} indicate the luminosity weighted temperature, \lbrat, and mass-weighted temperature, \mbrat, respectively, at the finest resolution. By construction, the black point lies on the blue line at the finest resolution. As we blur the galaxy towards a single resolution element, the black points should approach the red line. The difference between the red and blue lines shows the maximum effect of interchanging luminosity and mass weighting. This difference is $\sim 2{-}3$~K, or about $10\%$ change in $T_d$, or a factor of $\sim 2$ change in $U$.
% There is no difference between mass and luminosity weighted averages for a single data point, and the convolution is luminosity weighted.

{\bf Implication for the total mass measurement:} Luminosity weighting tends to bias towards higher temperature. An overestimate of temperature will lead to an underestimate of the mass. To see this, consider a point on the Rayleigh-Jeans tail, for which $I_\nu \propto T_d~\Sigma_d$. As the luminosity-weighting biases toward higher temperature, it will also lead to an underestimate of the mass.

To first order, the mass will be biased low by the same fraction as the temperature is biased high. This is $\sim 10{-}20\%$ change over the range of resolutions that we study. However, this does not capture the effects of any distributions of mass or temperature beneath our finest resolution (13~pc for Magellanic Clouds and 167~pc for M31 and M33).
% This only reflects the bias over the range of resolutions that we study.

For the LMC, \citet{galliano11} considered exactly this effect and estimated that the mass inferred from the integrated SED is biased low by $50\%$. They included shorter wavelength bands than what we study here, which likely accounts for that stronger bias than what we found. The effect of luminosity weighting should be stronger for these shorter wavelength bands because they are more sensitive to the dust temperature.

In general, following \citet{dale01}, accounting for distributions of temperatures in modeling the SED is crucial for a correct mass estimation. Emphasizing longer wavelengths, as we do here, somewhat diminishes this resolution effect. Even at 13~pc resolution, mass estimations in SMC and LMC are still less accurate because a significant portion of blending populations (unresolved by our study) must still occur even below that resolution.

\subsection{Fits to the Integrated SED and Comparison with Resolved Mapping}
\label{sec:int_fit}

We also fit the integrated SED, equivalent to unresolved observation. Even in the local universe, this is often the only type of measurement available for a galaxy. In particular, poor resolution options remain the main way to study high redshift galaxies \citep[e.g.,][]{magdis10,magdis12,genzel15} and there has been some controversy about how to best estimate dust mass in these galaxies. Here, following \citet{galliano11}, we check how the temperature and dust mass derived from the integrated SED compared to those derived from the highly resolved maps.

\begin{figure*}
\centering
\includegraphics[width=0.8\textwidth]{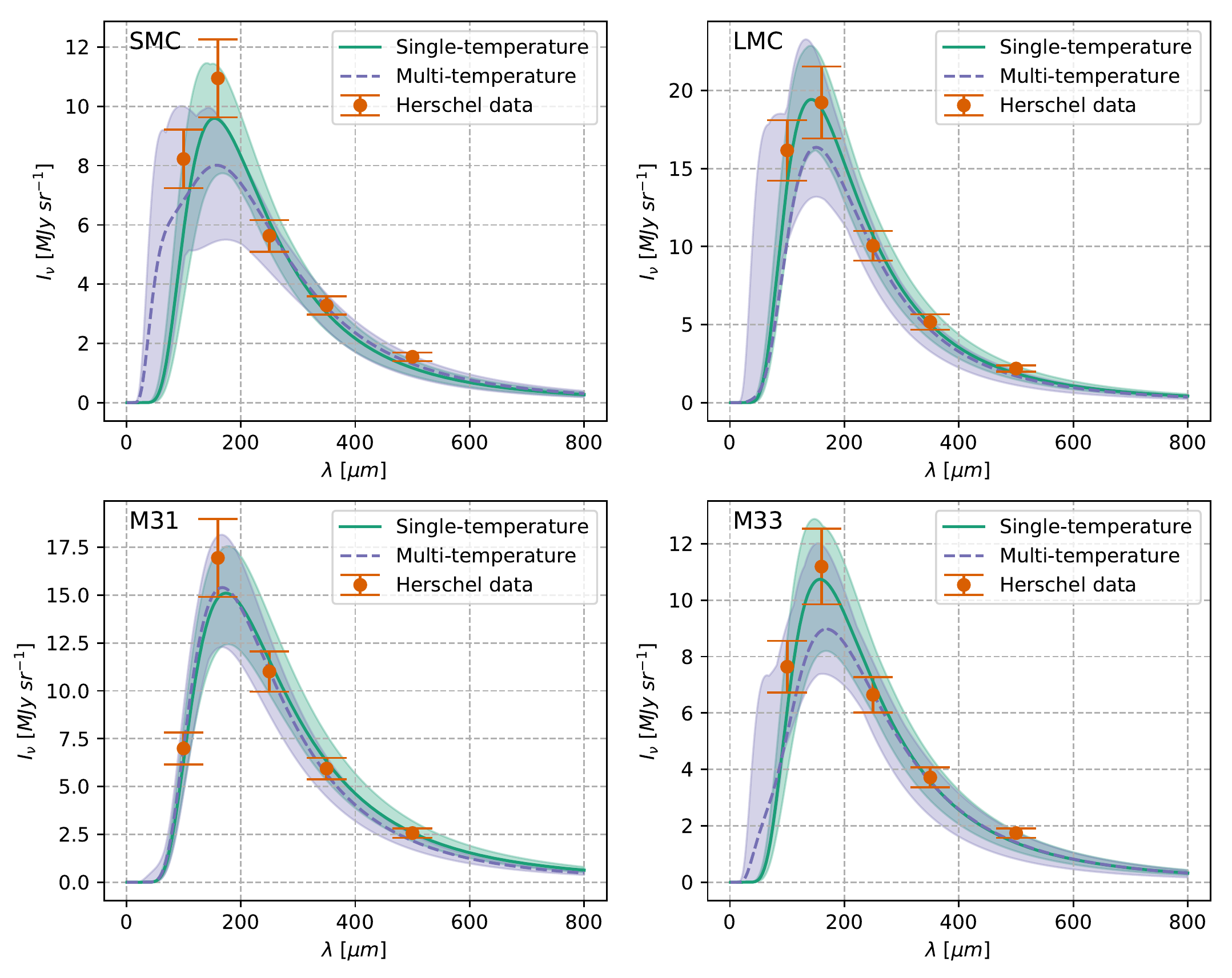}
\caption{{\bf Modified blackbody fits to the integrated far-infrared fluxes of each galaxy using a fixed $\beta = 1.8$.} The red points are the observed integrated SED, while the expected likelihood models for a single-temperature and multi-temperatures are shown as solid green and dashed purple curves, respectively. The shaded area are the minimum and maximum values of 100 models, draw with respect to their likelihood.}
\label{fig:int_sed}
\end{figure*}

We calculate the integrated SED by taking the mean surface brightness within the region of interest. We report the mean intensity for each target at each band in Table \ref{tab:summary}. We also note the averaging area, in steradians. To calculate the integrated SED in units of specific flux, we multiply the area by the intensities. Then, we fit the SED with two approaches; single-temperature and multi-temperature modified blackbodies. Figure~\ref{fig:int_sed} shows these integrated SEDs as red points, along with the best-fit models (curves). Uncertainty in the best-fit models, visualized by drawing 100 realizations in proportion to their likelihood, are shown as shaded areas.

In Table~\ref{tab:summary}, we compare the dust temperature and total mass between galaxies from three different approaches; (1) integrated SED, fitted using single-temperature modified blackbody, (2) integrated SED, fitted using multi-temperature modified blackbody, and (3) the mass-weighted mean of temperature, \mbrat, calculated from the finest resolution map.

{\bf Single-Temperature Model:} First, we use the same single-temperature modified blackbody model as adopted for the resolved observations. The best-fit models are shown as the solid green curves in Figure~\ref{fig:int_sed}. Despite the simplicity of our model, it offers a good fit to the integrated SED of each galaxy. Mathematically, this is different than the luminosity-weighted mean of temperature, \lbrat, because in the integrated SED, we take average of intensity without any weighting. Therefore, we do not expect \lbrat\ to be the same as the temperature derived from the integrated SED fitting (red vs. green lines in Figure~\ref{fig:temp_res}).

\begin{deluxetable*}{ c | c c c c c | c c c | c c c }
\tablewidth{0pt}
\tabletypesize{\scriptsize}
\tablecaption{Summary of measurements for the whole galaxy. \label{tab:summary}}
\tablehead{
{} & \multicolumn{5}{c |}{} & \multicolumn{3}{c |}{} & \multicolumn{3}{c}{} \\
Galaxies & \multicolumn{5}{c |}{Flux Density [$10^3$ Jy]} & \multicolumn{3}{c |}{Dust Equilibrium Temperature [K]} & \multicolumn{3}{c}{Total Dust Mass [$10^5~M_\odot$]} \\
{} & 100 \micron\ & 160 \micron\ & 250 \micron\ & 350 \micron\ & 500 \micron\ & Single \temp\ & Multi \temp\ & Resolved & Single \temp\ & Multi \temp\ & Resolved
}
\startdata
SMC & $14.07 \pm 1.69$ & $18.73 \pm 2.25$ & $9.64 \pm 0.92$ & $5.61 \pm 0.53$ & $2.65 \pm 0.25$ & $19.6_{1.2}^{0.7}$ & $17.6_{-1.0}^{-1.2}$ & $18.2 \pm 1.7$ & $1.6_{0.3}^{0.3}$ & $3.9_{1.1}^{1.3}$ & $1.9 \pm 0.7$ \\
LMC & $222.68 \pm 26.72$ & $264.83 \pm 31.78$ & $138.42 \pm 13.15$ & $71.19 \pm 6.76$ & $30.13 \pm 2.86$ & $21.3_{1.1}^{0.6}$ & $21.2_{-1.8}^{-1.7}$ & $20.0 \pm 1.3$ & $11.7_{1.9}^{1.5}$ & $15.9_{5.2}^{5.9}$ & $13.1 \pm 3.5$ \\
M31 & $3.59 \pm 0.43$ & $8.71 \pm 1.05$ & $5.66 \pm 0.54$ & $3.05 \pm 0.29$ & $1.32 \pm 0.13$ & $17.4_{0.7}^{0.1}$ & $18.1_{-0.9}^{-0.7}$ & $16.9 \pm 1.1$ & $195.7_{26.2}^{17.9}$ & $192.1_{22.0}^{8.7}$ & $195.0 \pm 46.5$ \\
M33 & $1.46 \pm 0.18$ & $2.14 \pm 0.26$ & $1.27 \pm 0.12$ & $0.71 \pm 0.07$ & $0.33 \pm 0.03$ & $19.2_{1.0}^{0.5}$ & $18.4_{-1.6}^{-1.4}$ & $16.3 \pm 1.3$ & $41.0_{7.1}^{5.9}$ & $73.2_{21.2}^{25.5}$ & $61.7 \pm 23.3$ \\
\enddata
\tablecomments{For multi-temperature model, we use parameters $U_{\rm min}$, $U_{\rm max}$, and $\gamma$ to derive the mean ISRF, $\langle U \rangle$, using Equation~\ref{eq:ubar}, and then convert it to dust temperature using Equation~\ref{eq:u2t}.}
\end{deluxetable*}

The expected value of \temp\ from fitting these integrated fluxes are shown as the green lines in Figure~\ref{fig:temp_res}. Dust temperature from fitting the integrated SED is higher than the mass-weighted temperature of the most highly resolved observations. This means the unresolved observations would overestimate the temperature of the majority of the dust by $\lesssim 3$~K or a factor of $\lesssim 17\%$ for 18~K temperature. If using a single flux at 500 \micron\ to measure the dust mass \citep[Equation~1 in][]{scoville14}, this translates to an underestimate of dust mass also by $\lesssim 17\%$ for an object with 18~K dust temperature.

{\bf Multi-temperature Model:} We also fit the integrated SED by adopting a sub-resolution multi-temperature distribution in the form of \citep[for $\alpha \neq 1$;][]{draine07a}
\begin{equation}
\label{eq:distU}
\frac{dm_d}{dU} = (1-\gamma)~\delta(U-U_{\rm min}) + \gamma \frac{(\alpha-1)}{U_{\rm min}^{1-\alpha} - U_{\rm max}^{1-\alpha}} U^{-\alpha}~,
\end{equation}
where $\delta$ is Dirac's delta function. Here, a fraction $\gamma$ of the dust mass within a resolution element is heated by a power law distribution field $U^{-\alpha}$ between $U_{\rm min}$ and $U_{\rm max}$, while the rest is heated by the ambient ISRF of $U_{\rm min}$.

We use the same grid-based fitting procedure as before (\S \ref{sec:fitting}). The grid for $\alpha$ is from $1.1$ to $5.1$ with an increment of $0.2$, the grid for log$_{10}\gamma$ is from $-3$ to $0$ with an increment of $0.2$, and the grid for log$_{10}U_{\rm min}$ from $-1.5$ to $1.5$ with an increment of $0.2$. The grids for dust temperature and mass surface density are the same as in the single temperature modified blackbody model (\S \ref{sec:fitting}). The value of $U_{\rm max}$ is fixed at $10^3$. The adoption of this $U_{\rm max}$ value does not affect the outcome of fitting parameters.

The best-fit models are shown as the dashed purple curves in Figure~\ref{fig:int_sed}. The expected values for $U_{\rm min}$, $\alpha$, and $\gamma$ are listed in Table~\ref{tab:params3}. The value of $U_{\rm min}$ represents the ambient ISRF that illuminate the majority ($1-\gamma$) of dust mass. Hence, it is not surprising that it is highest in the LMC, in general agreement with the distribution of dust mass as a function of $U$ ($\S$\ref{sec:dis_mass_by_u}). The rest of dust mass (a fraction of $\gamma$) is heated by ISRF in the range between $U_{\rm min}$ and $U_{\rm max}$.

\begin{deluxetable}{ c | c c c }
\tablewidth{0pt}
\tabletypesize{\scriptsize}
\tablecaption{The expected values of integrated SED multi-temperature fits. \label{tab:params3}}
\tablehead{
\\
Galaxies & $\alpha$ & log$_{10} U_{\rm min}$ & $\gamma$ \\
{} & {} & [dex] & [\%]
}
\startdata
%SMC & $1.61_{-0.44}^{+0.19}$ & $-0.53_{-0.12}^{+0.31}$ & $20.75_{-16.17}^{+26.86}$ \\
%LMC & $2.15_{-0.96}^{+1.02}$ & $0.12_{-0.18}^{+0.38}$ & $8.68_{-7.44}^{+29.35}$ \\
%M31 & $3.31_{-1.29}^{+1.08}$ & $-0.11_{-0.03}^{+0.17}$ & $1.50_{-1.35}^{+8.84}$ \\
%M33 & $1.82_{-0.64}^{+0.21}$ & $-0.29_{-0.14}^{+0.34}$ & $13.32_{-10.80}^{+29.85}$ \\
SMC & $1.61_{-0.45}^{+0.19}$ & $-0.51_{-0.32}^{+0.12}$ & $20.73_{-16.16}^{+26.80}$ \\
LMC & $2.15_{-0.95}^{+1.01}$ & $0.12_{-0.38}^{+0.18}$ & $8.78_{-7.50}^{+29.36}$ \\
M31 & $3.31_{-1.29}^{+1.08}$ & $-0.11_{-0.17}^{+0.03}$ & $1.51_{-1.36}^{+8.88}$ \\
M33 & $1.77_{-0.60}^{+0.20}$ & $-0.32_{-0.33}^{+0.13}$ & $14.27_{-11.49}^{+29.65}$ \\
\enddata
\end{deluxetable}

The value of $\alpha$ shows the relative contribution of dust mass at the high-end tail of $U$, compared to that in the low-end tail of $U$, where $U_{\rm min} < U < U_{\rm max}$. A shallower slope (lower $\alpha$), such as in the SMC and M33, means the low-end tail of $U$ (originated from star-forming regions) contributed more to the dust mass spectrum between $U_{\rm min}$ and $U_{\rm max}$. Hence, this multi-temperature fit is not only adding free parameters, but also physically more reasonable than a single-temperature fit of modified blackbody.

Following \citet{draine07a}, we calculate the mass-weighted mean of $U$ via
\begin{equation}
\label{eq:ubar}
\langle U \rangle_M = (1-\gamma)U_{\rm min} + \gamma \frac{U_{\rm min} \ {\rm ln}(U_{\rm max}/U_{\rm min})}{1-U_{\rm min}/U_{\rm max}}~.
\end{equation}
Then, we convert $\langle U \rangle_M$ to the dust temperature by inverting Equation~\ref{eq:u2t}.

{\bf Comparison between methods:} In SMC and M33, the temperature derived from integrated SED multi-temperature is lower than those derived from the integrated SED single-temperature model by $\sim1-2$~K, while the opposite happens in M31. The best-fit temperature from two methods above matches in LMC. Compared to \mbrat, the temperature derived from integrated SED (using both single- and multi-temperature model) has higher temperature than \mbrat\ by $\gtrsim 1$~K (except for integrated multi-temperature in the SMC). This difference is because the integrated SED convolves together many regions so it is biased toward warm regions (for single-temperature model), and the limitation in modeling the integrated SED to recover the true distribution of $dM_d/dU$ as seen in the highly resolved map (for multi-temperature model).

For the total dust mass, all three methods (integrated SED with single-temperature, integrated SED with multi-temperature, and by summing mass from all pixels in the finest resolution map) agree remarkably well in M31. In the Magellanic Clouds and M33, the mass from the finest resolution map is in between results from the single- and multi-temperature methods. In the Magellanic Clouds, the single-temperature method gives better agreement (but with lower mass) compared to the total mass from resolved map, while in M33, we find better agreement (but with higher mass) between the total masses from multi-temperature method and the resolved map.

\section{Summary \label{sec:sum}}

Using the archival {\em Herschel} maps covering from $\lambda = 100$ to $500$~\micron\ and a fitting algorithm developed by \citet{gordon14} and implemented by \citet{chiang18}, we derive maps of equilibrium dust temperature ($T_d$), dust mass surface density ($\Sigma_d$), and equilibrium infrared luminosity ($\mathcal{J_{\rm IR}}$) for four Local Group galaxies: the Small and Large Magellanic Cloud, M31, and M33. We show these maps, which we make publicly available, in Figure~\ref{fig:maps_temp}. We construct the maps at $13$~pc resolution, the common physical resolution available for the Magellanic Clouds, and $167$~pc resolution, the common physical resolution available for all four targets.

We use these maps to measure how the dust mass and luminosity are distributed as functions of $T_d$ (or equivalently, the interstellar radiation field strength, $U \propto T_d^{5.8}$), and $\Sigma_d$, in each target. We show these distributions in Figures~\ref{fig:dist_13pc} and~\ref{fig:dist_167pc}. 

Based on these calculations, we gauge the median dust temperatures and surface densities by mass and luminosity and note the characteristic widths of these distributions. We report these values in Table \ref{tab:params}. We highlight the following key points.

\begin{enumerate}
\item At $167$~pc resolution, the distribution of dust mass as a function of radiation field implies a median $T_d \approx 16{-}20$~K. If we instead consider the distribution of luminosity, the median $T_d$ appears higher, $T_d \approx 18{-}21$~K. On average, the peak of $T_d$ weighting by luminosity is only $\sim 1$~K higher than the peak of $T_d$ weighting by mass, but in M33, this shift is about $6$~K, reflecting the strong radial structure in that galaxy.

\item At $167$~pc resolution, $68\%$ of the dust mass is spread over $\sim 0.4{-}1$~dex in radiation field, $U$. The luminosity spans a lower range, $\sim 0.4{-}0.8$~dex. Again, the strong radial structure and extended disk in M33 lead to a wider distribution, $\sim 0.8{-}1$~dex, than we see in the other galaxies ($\sim 0.4{-}0.6$~dex).

\item Also at $167$~pc resolution, the dust mass as a function of surface density (i.e., the dust mass surface density PDF) has median value $\log_{10} \Sigma_{\rm d} \approx -1.34$ to $-1.65$, i.e., $\Sigma_d \approx 0.02{-}0.05$~M$_\odot$~pc$^{-2}$. This agrees well with observation that the neutral ISM in all of our targets is mostly {\sc Hi} with dust to gas ratios of a few hundred to a thousand. The median $\Sigma_d$ for the distribution of luminosity is usually $\sim 0.1$~dex lower than the distribution by mass.

\item The widths of the distributions of both mass and luminosity as a function of $\Sigma_d$ are $\approx 0.5{-}0.7$~dex. This agrees well with the width measured for gas (atomic and molecular) column density PDFs \citep[e.g.,][]{wada07,berkhuijsen15,corbelli18,sun18}.
\end{enumerate}

Both the dust mass and luminosity PDFs show significant structure. This structure relates to features visible in the temperature, luminosity, and surface density maps. For example, M33's strong radial temperature gradient leads to a wide, multi-component distribution in dust mass as a function of temperature. Similarly, M31's star-forming ring and hot but low-surface density bulge region manifest as clear features in the distribution.

To further quantify how regional features contribute to the global distribution, we identify the active star-forming part of each galaxy and construct a separate distribution for this region and the quiescent region (Figures~\ref{fig:no_pdr_13} and \ref{fig:no_pdr_167}). We find the following.

\begin{enumerate} \setcounter{enumi}{4}
\item As expected, the star-forming parts of galaxies have warmer dust temperature and higher dust mass surface density than the rest of the regions in the galaxy. In our sample, the notable exception is M31, where dust in the central region is heated by older stellar populations \citep{groves12,smith12}. This dust stands out in our sample because of its combination of low mass surface density but high dust temperature.

\item Although these star-forming regions contribute a small fraction to the total dust mass ($10-28\%$), they produce a significant fraction of the equilibrium luminosity in the infrared ($19-51\%$).
\end{enumerate}

Furthermore, we see a more general form of these results by directly plotting the correlation between $\Sigma_d$ and $T_d$ for the points with high $S/N$ (Figure~\ref{fig:mass_temp}). We find the following.

\begin{enumerate}\setcounter{enumi}{6}
\item Dust temperature ($T_d$) correlates with dust mass surface density ($\Sigma_d$) in SMC, LMC, and M31 (Spearman rank correlation coefficients are weak, but highly significant). In M31, a weak correlation is evident in the dust associated with the star-forming ring, but the hot, low surface density bulge dust heated by the old stellar population leads to an overall anti-correlation between $T_d$ and $\Sigma_d$.

\item Very roughly, the sense of the correlation between $\Sigma_d$ and $T_d$ agrees with expectations from an approximately quadratic gas-star formation scaling relation ($\Sigma_{\rm SFR} \propto \Sigma_{\rm gas}^2$), as might be roughly expected in the {\sc Hi}-dominated regime appropriate for our targets.
\end{enumerate}

The \textit{Herschel} Space Observatory observed the Local Group galaxies at high physical resolutions. We compare our results to those expected for the same galaxies at farther distances (higher redshift) by progressively convolving the galaxies to coarser resolutions. We find the following.

\begin{enumerate} \setcounter{enumi}{8}
\item The luminosity-weighted mean temperature typically exceeds the mass-weighted mean temperature, \mbrat, by $\sim 2{-}3$~K, an increase of about 10\% (or about a factor of $\sim 2$ in $U$). Progressively blurring the maps with a luminosity weighting causes the low resolution \mbrat\ to increase, approaching this higher value (Figure~\ref{fig:temp_res}).

\item The best-fit temperature derived from the integrated SED (equivalent to unresolved source) is higher than \mbrat\ from the finest resolution map (Table~\ref{tab:summary}). This bias is somewhat reduced when a sub-resolution multi-temperature distribution \citep[e.g., the approach in][]{dale01,draine07a} is used.

\item Both points above quantify how severely the temperature from low resolution IR-SED fitting in distant galaxies may overestimate their true mass-weighted mean temperature. As a consequence, the inferred total dust mass in distant galaxies would be underestimated. This is particularly important when the gas mass is determined based on the dust mass.
\end{enumerate}

Lastly, we compare our five-bands fitting results to those that would be obtained by using only one or two bands, as is now commonly required with ALMA \citep{scoville14}. We test how the 500 \micron\ intensity, $I_{500\micron}$, performs as a tracer for dust mass surface density, $\Sigma_d$. We also test how well our best-fit $T_d$ matches individual infrared color ($I_{100\micron}/I_{250\micron}$ or $I_{100\micron}/I_{350\micron}$). We find the following.

\begin{enumerate} \setcounter{enumi}{10}
\item A median radio of $I_{500\micron}$ over $\Sigma_d$ can recover the trend between $I_{500\micron}$ and $\Sigma_d$ in our data (Figure~\ref{fig:corrs}). There are two subtleties. First, there is $\lesssim 50\%$ offset in this median ratio from galaxy-to-galaxy. Second, real temperature variations, e.g., as seen in M33 gradient, lead to systematic bias and a modestly sublinear slope between $\Sigma_d$ and $T_d$. The sense is that without accounting for radial trend in dust temperature, one would estimate that dust in M33 to be more confined to the galaxy center than it actually is.

\item Although it is an implication of our modeling method to determine the dust temperature, the modified black-body model with a single-temperature and a constant emissivity index, $\beta$, can adequately relate the infrared color and dust temperature (Figure~\ref{fig:temp_color}), provided that the wavelengths (used as the IR-color) bracket the peak wavelength of the cold dust emission. We observe scatter in this color--temperature relation, particularly in the low temperature. Also, the color--temperature relation in the SMC looks steeper than the prediction from the model, probably due to variation of $\beta$ from region-to-region \citep{gordon14}.
\end{enumerate}

We make the maps of dust temperature and mass surface density in the {\tt FITS} file to be publicly available at \url{https://www.asc.ohio-state.edu/astronomy/dustmaps}.

\acknowledgments
We thank the referee for thoughtful comments that improved the paper. DU, IC, AKL, KMS, and JC are supported by the National Science Foundation (NSF) Grant No. 1615728 and NASA ADAP grants NNX16AF48G and NNX17AF39G. DU and AKL are also partially supported by NSF under Grants No. 1615105, 1615109, and 1653300. {\it Herschel} is an ESA space observatory with science instruments provided by European-led Principal Investigator consortia and with important participation from NASA. This work is based, in part, on observations made with the {\it Spitzer} Space Telescope, which is operated by the Jet Propulsion Laboratory, California Institute of Technology under a contract with NASA. Partial support was also provided to AKL in the context of HST project HST-GO-13659.013-A (PI Sandstrom), which is studying dust in the Magellanic Clouds. That project is based on observations made with the NASA/ESA Hubble Space Telescope, obtained from the data archive at the Space Telescope Science Institute. STScI is operated by the Association of Universities for Research in Astronomy, Inc. under NASA contract NAS 5-26555.

\facilities{{\it Herschel}(PACS and SPIRE), {\it Spitzer}(MIPS).}
\software{CASA \citep[v4.7.2; ][]{mcmullin07}.}

\appendix

\section{Challenge in comparing across physical resolutions\label{app: cov}}

One challenge in fitting the dust properties across physical resolutions is computing the covariance matrix of the background fluxes ($\S$\ref{sec:fitting}) for the low-resolution images. The convolution process blurs the image.  Therefore, to conserve flux from a galaxy, the background area decreases and the region of interest area increases in the low resolution maps. This means we can only compute the covariance matrix of background flux correctly down to certain resolution. In this study, this lower limit in resolution is selected to be where there are at least ten resolution elements across the target image.

It is shown in \citet{chiang18} that the choice of covariance matrix heavily affects the results of fitting. Thus without the covariance matrix of background fluxes, we are unable to fit dust properties at coarser resolutions. Several attempts have been made to extrapolate the covariance matrix to coarser resolutions. So far, none of the methods we try are proved to give the correct matrix mainly due to two reasons. (1) The background flux consists of both random noise and real signal from background objects. (2) The observed correlations between \textit{Herschel} bands are not constant across resolutions. Whether it is possible to extrapolate the covariance matrix for low-resolution images from high-resolution images remains an open question for future study.

\bibliographystyle{yahapj}
\bibliography{references}

\end{document}